\documentclass[twocolumn]{aastex701}

\begin{document}

\title{CHANG-ES {XLII}: Cosmic-Ray Electron Transport and the Spatially Resolved Radio--SFR Relation in Edge-on Galaxies}

\author[0009-0006-3887-8988, gname=Jianghui, sname=Xu]{Jianghui Xu}
\affiliation{Department of Astronomy, University of Science and Technology of China, Hefei, Anhui 230026, People's Republic of China}
\affiliation{School of Astronomy and Space Science, University of Science and Technology of China, Hefei 230026, People's Republic of China}
\affiliation{Hamburger Sternwarte, Universit\"at Hamburg, Gojenbergsweg 112, D-21029 Hamburg, Germany}
\email{jianghuixu@mail.ustc.edu.cn}

\author[0000-0002-2082-407X, gname=Volker, sname=Heesen]{Volker Heesen}
\affiliation{Hamburger Sternwarte, Universit\"at Hamburg, Gojenbergsweg 112, D-21029 Hamburg, Germany}
\email{volker.heesen@uni-hamburg.de}

\author[0000-0001-6239-3821, gname=Jiang-Tao, sname=Li]{Jiang-Tao Li}
\affiliation{Purple Mountain Observatory, Chinese Academy of Sciences, 10 Yuanhua Road, Nanjing 210023, China}
\email[show]{pandataotao@gmail.com}

\author[0000-0003-4286-5187, gname=Guilin, sname=Liu]{Guilin Liu}
\affiliation{Department of Astronomy, University of Science and Technology of China, Hefei, Anhui 230026, People's Republic of China}
\affiliation{School of Astronomy and Space Science, University of Science and Technology of China, Hefei 230026, People's Republic of China}
\email[show]{glliu@ustc.edu.cn}

\author[0009-0001-8154-3562, gname=Rainer, sname=Beck]{Rainer Beck}
\affiliation{Max-Planck-Institut f\"ur Radioastronomie, Auf dem H\"ugel 69, 53121 Bonn, Germany}
\email{rbeck@mpifr-bonn.mpg.de}

\author[0000-0002-3286-5346, gname=Li-Yuan, sname=Lu]{Li-Yuan Lu}
\affiliation{Department of Physics and Astronomy \& Research Center of Astronomy, Qinghai University, 251 Ningda Road, Xining 810016, People’s Republic of China}
\email{lylu@qhu.edu.cn}

\author[0000-0001-7254-219X, gname=Yang, sname=Yang]{Yang Yang}
\affiliation{Xiangtan University, Xiangtan 411105, Hunan, People’s Republic of China}
\email{yangyang.astro@gmail.com}

\author[0000-0001-5310-1022, gname=Jayanne, sname=English]{Jayanne English}
\affiliation{Department of Physics \& Astronomy, University of Manitoba, Winnipeg, Manitoba, R3T 2N2, Canada}
\email{jayanne.english@umanitoba.ca}

\author[0009-0003-9052-1976, gname=Shengtao, sname=Wang]{Shengtao Wang}\affiliation{School of Physics and Astronomy, Yunnan University, Kunming 650500, China}
\email{wstfch@mail.ynu.edu.cn}

% \author{Co-Authors}
% \affiliation{Affiliation of Co-Authors}
% \email{email of Co-Authors}

\begin{abstract}

Radio continuum emission traces star formation in galaxies, but cosmic-ray electron (CRE) propagation between the disk and halo complicates this connection. We present a wide-band, spatially resolved analysis of 19 nearby edge-on galaxies, combining LOFAR observations at 144~MHz with CHANG-ES VLA imaging at 1.575, 3.0, and 6.0~GHz, and hybrid H$\alpha$+22~$\mu$m star formation rate (SFR) maps. All maps are convolved to a common $2.1\,{\rm kpc}$ resolution for pixel-by-pixel analysis of the nonthermal spectral index $\alpha_{\rm nth}$ and the SFR surface density $\Sigma_{\rm SFR}$. All 19 galaxies exhibit statistically significant nonthermal spectral flattening with increasing $\Sigma_{\rm SFR}$, with a sample median slope of $0.19_{-0.03}^{+0.09}$ that traces the contrast between CRE injection in the star-forming disk and cooling in the halo. The galaxy-to-galaxy scatter in this slope shows no dependence on global galaxy properties and instead appears to be modulated by the local dynamical environment, with the largest $\alpha_{\rm nth}$--$\Sigma_{\rm SFR}$ slopes appearing in tidally perturbed systems. The resolved radio--SFR slope steepens from $0.57$ at 144~MHz to $0.95$ at 6~GHz, approaching linearity at high frequencies. This trend is consistent with energy-dependent CRE cooling and transport, with the deviation from linearity most pronounced at low frequencies, where the edge-on line of sight blends midplane and halo CRE populations. These results are robust against thermal contamination, beam smearing, and spectral curvature. Together, our analysis establishes $\alpha_{\rm nth}$ as an observational probe of projected disk-to-halo CRE transport in edge-on galaxies and quantifies how this transport reshapes the frequency dependence of resolved radio--SFR calibrations.

\end{abstract}


\section{Introduction}\label{sec:introduction}

The tight empirical correlation between far-infrared and radio continuum emission in galaxies has long established radio observations as a crucial, extinction-free tracer of star formation \citep{Condon02}. In star-forming galaxies, the radio continuum is composed of two components: thermal free-free emission, which traces ionizing photons from young massive stars and thus the instantaneous star formation rate (SFR), and nonthermal synchrotron emission, which is produced by cosmic-ray electrons (CREs) accelerated in supernova remnants and gyrating in the interstellar magnetic field. At frequencies $\lesssim 10\,{\rm GHz}$, the nonthermal component dominates, typically accounting for $\gtrsim 80\%$ of the total flux density \citep{Tabatabaei17}. A strictly linear correlation between the SFR and radio continuum emission is expected if the CRE production rate traces the supernova rate, and if these electrons radiate away nearly all their energy via synchrotron and inverse-Compton (IC) processes before escaping the host galaxy \citep{Condon92}. In reality, however, CREs propagate through the multi-phase interstellar medium (ISM) over kiloparsec scales before cooling, escaping into the halo, or being advected outward by galactic winds \citep[e.g.,][]{Heesen18,Krause18}. These transport processes lead to deviations from a purely linear radio--SFR relation \citep{Li16,Smith21,Tabatabaei17,Tabatabaei22,Heesen24}. How CREs propagate and lose energy is therefore central to interpreting radio continuum emission as a quantitative tracer of local star formation.

The nonthermal spectral index, $\alpha_{\rm nth}$, provides a direct diagnostic of this propagation and cooling history. Freshly injected CREs from supernova remnants exhibit a flat power-law spectrum with an injection index of $\alpha_{\rm nth} \sim -0.5$ (for $I_{\nu} \propto \nu^{\alpha_{\rm nth}}$), set by diffusive shock acceleration \citep{Bell78,Reynolds08}. As CREs propagate away from their injection sites, they lose energy through synchrotron and IC cooling with a characteristic cooling timescale of $t_{\rm cool} \propto E^{-1} \propto \nu^{-1/2}$ \citep{Condon92}, causing the synchrotron spectrum to progressively steepen toward $\alpha_{\rm nth} \sim -1.0$. Consequently, spatially resolved $\alpha_{\rm nth}$ maps trace the balance between fresh acceleration and aging within a galaxy: flat indices pinpoint active injection sites, while steep indices trace regions dominated by cooled, propagated electrons. Both global and resolved studies have observationally corroborated this picture. \citet{Tabatabaei17} reported a global scaling relation between $\alpha_{\rm nth}$ and the SFR surface density ($\Sigma_{\rm SFR}$) with a slope of $\sim 0.17$ across nearby galaxies, and highly resolved observations of M~33 yielded a comparable slope of $\sim 0.21$ \citep{Tabatabaei22}. However, these measurements are intrinsically limited to characterizing the in-plane diffusion of CREs across the galactic disk. In a face-on view, the star-forming midplane and the extended synchrotron halo are superimposed along the line of sight, rendering the vertical CRE transport that physically connects them observationally inaccessible.

Edge-on galaxies offer the geometry required to disentangle this vertical structure. When viewed at high inclinations, the star-forming midplane is spatially separated from the extraplanar synchrotron halo, allowing the injection-dominated and cooling-dominated regimes of the CRE population to be probed as distinct spatial environments. The Continuum HAlos in Nearby Galaxies---an EVLA Survey \citep[CHANG-ES;][]{Irwin12,Irwin12b} has demonstrated that radio halos extending several kiloparsecs above the midplane are nearly ubiquitous in star-forming edge-on systems. Their characteristic scale heights and vertical transport timescales vary systematically depending on the galactic environment \citep{Krause18,Mora-Partiarroyo19a,Stein19,Stein23,Heesen25,Xu25,Xu26}. A complementary observational probe of CRE transport is provided by the frequency dependence of the spatially resolved radio--SFR relation. Because the synchrotron cooling time decreases with increasing frequency, high-frequency emission is dominated by short-lived, energetic CREs that are still observed close to their injection sites, whereas low-frequency emission traces older, lower-energy electrons that have diffused outward into the halo \citep{Mulcahy14,Heesen19,Heesen21}. The resolved radio--SFR slope is therefore expected to flatten toward lower frequencies, with the magnitude of this gradient directly tracing the energy-dependent CRE diffusion length scale. Indeed, the spatially resolved 144~MHz to 1.4~GHz study by \citet{Heesen24} found flatter slopes at lower frequencies in a predominantly face-on sample.

So far, however, the diagnostic power of $\alpha_{\rm nth}$ and the geometric advantages of edge-on systems have largely been exploited in isolation. Spatially resolved $\alpha_{\rm nth}$--$\Sigma_{\rm SFR}$ and radio--SFR scaling relations have been developed almost exclusively using face-on or low-inclination samples \citep{Tabatabaei17,Tabatabaei22,Heesen24}, while edge-on radio studies have, in turn, focused largely on the vertical structure of individual halos rather than on resolved scaling relations \citep{Schmidt19,Miskolczi19,Stein19}. A comprehensive, wide-band statistical analysis uniting these two threads remains lacking. By tracking both the spatially resolved $\alpha_{\rm nth}$ diagnostic and the frequency-dependent radio--SFR slope in an edge-on sample, joint observational tracers of disk-to-halo CRE transport can be established. Such an analysis offers a path toward a more complete physical picture of how CREs are injected, transported, and cooled across the full vertical extent of star-forming disks, while also clarifying the extent to which inclination-driven projection effects shape radio-based SFR tracers. The latter is of practical relevance to high-redshift studies, where star-forming galaxies are routinely observed at low frequencies with random inclinations and limited spatial resolution. 

To address these questions, we present a spatially resolved, multi-band analysis of 19 edge-on galaxies, combining LOFAR observations at 144~MHz with Karl G. Jansky Very Large Array (VLA) imaging at 1.575, 3.0, and 6.0~GHz, together with hybrid H$\alpha$+22~$\mu$m SFR maps. All maps are matched to a common physical resolution of $2.1\,{\rm kpc}$, enabling a uniform pixel-by-pixel analysis of the nonthermal spectral index $\alpha_{\rm nth}$ and the SFR surface density $\Sigma_{\rm SFR}$. We use this dataset to characterize the spatially resolved $\alpha_{\rm nth}$--$\Sigma_{\rm SFR}$ relation as a local diagnostic of CRE injection versus cooling, to measure the frequency dependence of the radio--SFR slope as an independent probe of energy-dependent transport, and to interpret both in the framework of disk-to-halo CRE transport in highly inclined geometries. The paper is organized as follows. Section~\ref{sec:data} describes the sample, multi-wavelength data, and fitting methodology. Section~\ref{sec:result} presents the spatially resolved $\alpha_{\rm nth}$--$\Sigma_{\rm SFR}$ scaling and the frequency-dependent radio--SFR relations. Section~\ref{sec:discussion} interprets these findings within the context of CRE transport physics, examines inclination effects, and evaluates the robustness of our results against thermal contamination, beam smearing, and spectral curvature. Section~\ref{sec:summary} summarizes our primary conclusions.

\section{Data and Methodology}\label{sec:data}
\subsection{Sample and Multi-wavelength Data}\label{subsec:data}

In this study, we utilize radio continuum observations from LOFAR and VLA. The LOFAR data are drawn from the third data release of the LOFAR Two-metre Sky Survey \citep[LoTSS-DR3;][]{LoTSS,LoTSS_DR3} at 144~MHz, providing maps at both $6\arcsec$ and $20\arcsec$ resolutions. Our VLA dataset spans the $L$-, $S$-, and $C$-bands, originating from CHANG-ES \citep{Irwin12, Irwin12b}. The 1.575~GHz $L$-band and 6.0~GHz $C$-band data were acquired in the C and D configurations, respectively. Obtained from CHANG-ES Data Release 5 (R.~Walterbos et al.\ 2026, in preparation), these maps have been convolved to a common $15\arcsec$ resolution. Additionally, we include 3.0~GHz $S$-band observations from \citet{Heesen25}, which were taken in the C configuration with a resolution of $7\arcsec$. SFR maps were also obtained from CHANG-ES DR5 (R.~Walterbos et al.\ 2026, in preparation). These maps were derived from a combination of H$\alpha$ and WISE 22~$\mu$m emission, following the methodology described in \citet{Vargas19}. The input H$\alpha$ and WISE maps were matched to a common Gaussian beam with an FWHM of $15\arcsec$, using a Gaussian kernel for the H$\alpha$ maps and a kernel from \citet{Aniano11} for the WISE maps. The resulting SFR maps were therefore provided at a resolution of $15\arcsec$ (see Sections~4.2 and~4.3 of R.~Walterbos et al.\ 2026, in preparation, for details).

We identified 21 galaxies with available data across all the aforementioned datasets. To conduct a spatially resolved analysis, all maps must be convolved to a uniform physical resolution, which is inherently limited by the most distant galaxy in the sample. We excluded two galaxies, NGC~3735 ($d = 42.0\,{\rm Mpc}$) and NGC~5297 ($d = 40.4\,{\rm Mpc}$), whose distances place them significantly beyond the rest of the sample and would have degraded the common physical resolution to $\gtrsim 3\,{\rm kpc}$. Our final sample therefore consists of 19 galaxies, whose basic physical properties are summarized in Table~\ref{tab:sample_properties}. 

By utilizing the high-resolution LOFAR maps, we successfully convolved all images (LOFAR, $L/S/C$-bands, and SFR) across the sample to a common physical resolution of $2.1\,{\rm kpc}$, which corresponds to an angular resolution of $15\arcsec$ at the distance of NGC~5775, the most distant object in the final sample. For the quantitative analysis, all maps were then regridded to a pixel size equal to the beam FWHM ($2.1\,{\rm kpc}$), so that each pixel samples one independent resolution element and the pixel-by-pixel measurements are spatially independent. Exemplar multi-band maps are shown in Figure~\ref{fig:exemplar_maps} for NGC~891; for display clarity, these maps are presented at a finer pixel scale of $0.4\,{\rm kpc}$.

The noise level $\sigma_{\rm rms}$ for each map was estimated using the \texttt{sigma\_clipped\_stats} from the \texttt{Astropy} package, adopting a clipping threshold of $3.0\sigma$ and a maximum of 5 iterations. Each map was then masked at the $3\sigma_{\rm rms}$ level, and only pixels satisfying this threshold simultaneously in all maps were retained for subsequent analysis. This joint detection criterion excludes radio-only pixels without a hybrid SFR counterpart above the detection threshold, but does not remove diffuse radio emission within the retained pixels. No disk--halo decomposition or subtraction of diffuse nonthermal emission was performed. No dedicated mask was applied to AGN cores or associated extended emission, so these regions were retained wherever they satisfied the same selection criteria. The total uncertainty for each pixel was defined as $\sigma = \sqrt{\sigma_{\rm rms}^{2}+(\epsilon I)^{2}}$, where $I$ represents the pixel value (i.e., the radio continuum intensity or the SFR), and $\epsilon$ represents the adopted fractional uncertainty, representing calibration uncertainty for the radio maps and an approximate systematic uncertainty for the SFR maps. We adopt $\epsilon = 0.06$ for LOFAR \citep[following][]{LoTSS_DR3}, $\epsilon = 0.05$ for all VLA bands, and $\epsilon = 0.30$ for the SFR maps (R.~Walterbos et al.\ 2026, in preparation). 

\begin{deluxetable*}{lcccccccc}
\tablecaption{Basic Parameters of the Sample Galaxies\label{tab:sample_properties}}
\tablehead{
\colhead{Galaxy} & \colhead{Type} & \colhead{$d$} & \colhead{$M_*$} & \colhead{SFR} & \colhead{$\Sigma_{\rm SFR}$} & \colhead{$i$} & \colhead{PA} & \colhead{AGN} \\
\colhead{} & \colhead{} & \colhead{(Mpc)} & \colhead{($10^{10}\,M_{\odot}$)} & \colhead{($M_{\odot}\,\rm yr^{-1}$)} & \colhead{($10^{-3}\,M_{\odot}\,\rm yr^{-1}\,kpc^{-2}$)} & \colhead{($\degr$)} & \colhead{($\degr$)} & \colhead{}
}
\colnumbers
\startdata
NGC~891 & $3.1 \pm 0.4$ & 9.1 & $4.13 \pm 0.06$ & $1.89 \pm 0.18$ & $3.81 \pm 0.36$ & 84 & 22 & RF \\
NGC~2683 & $3.0 \pm 0.4$ & 6.27 & $1.49 \pm 0.02$ & $0.23 \pm 0.03$ & $3.32 \pm 0.43$ & 79 & 44 & RF \\
NGC~2820 & $5.3 \pm 0.6$ & 26.5 & $0.467 \pm 0.013$ & $1.06 \pm 0.11$ & $7.17 \pm 0.74$ & 88 & 65 & RF \\
NGC~3003 & $4.4 \pm 0.8$ & 25.4 & $0.485 \pm 0.010$ & $1.53 \pm 0.16$ & $2.54 \pm 0.27$ & 85 & 79 & RF \\
NGC~3044 & $5.5 \pm 0.7$ & 20.3 & $0.660 \pm 0.013$ & $1.71 \pm 0.16$ & $6.63 \pm 0.62$ & 85 & 113 & RF \\
NGC~3079 & $6.4 \pm 1.1$ & 20.6 & $4.73 \pm 0.07$ & $5.01 \pm 0.45$ & $9.48 \pm 0.85$ & 88 & 166 & ERB \\
NGC~3432 & $8.9 \pm 0.5$ & 9.42 & $0.100 \pm 0.002$ & $0.47 \pm 0.06$ & $6.05 \pm 0.77$ & 85 & 41 & RF \\
NGC~3448 & $3.8 \pm 4.2$ & 24.5 & $0.564 \pm 0.011$ & $1.74 \pm 0.18$ & $14.3 \pm 1.5$ & 78 & 65 & RF \\
NGC~3556 & $5.9 \pm 0.5$ & 14.09 & $2.81 \pm 0.04$ & $3.52 \pm 0.30$ & $7.22 \pm 0.62$ & 81 & 79 & RF \\
NGC~3628 & $3.1 \pm 0.3$ & 8.5 & $2.83 \pm 0.04$ & $1.26 \pm 0.11$ & $2.68 \pm 0.23$ & 87 & 104 & RB \\
NGC~4013 & $3.1 \pm 0.5$ & 16.0 & $3.23 \pm 0.05$ & $0.61 \pm 0.06$ & $3.01 \pm 0.30$ & 88 & 65 & RB \\
NGC~4096 & $5.3 \pm 0.6$ & 10.32 & $0.613 \pm 0.010$ & $0.56 \pm 0.07$ & $5.15 \pm 0.64$ & 82 & 20 & RF \\
NGC~4157 & $3.3 \pm 0.7$ & 15.6 & $2.92 \pm 0.04$ & $1.68 \pm 0.17$ & $7.72 \pm 0.78$ & 83 & 65 & RF \\
NGC~4217 & $3.0 \pm 0.4$ & 20.6 & $4.74 \pm 0.07$ & $1.76 \pm 0.17$ & $4.11 \pm 0.40$ & 86 & 50 & RF \\
NGC~4565 & $3.2 \pm 0.7$ & 11.9 & $6.04 \pm 0.08$ & $0.88 \pm 0.09$ & $0.86 \pm 0.09$ & 86 & 135 & RB \\
NGC~4631 & $6.5 \pm 0.7$ & 7.4 & $0.960 \pm 0.015$ & $2.25 \pm 0.20$ & $5.26 \pm 0.47$ & 85 & 86 & RF \\
NGC~4666 & $5.0 \pm 0.8$ & 27.5 & $12.48 \pm 0.18$ & $10.40 \pm 0.92$ & $12.8 \pm 1.1$ & 76 & 41 & RF \\
NGC~5775 & $5.1 \pm 0.7$ & 28.9 & $7.72 \pm 0.10$ & $7.17 \pm 0.62$ & $8.95 \pm 0.77$ & 86 & 145 & RF \\
NGC~5907 & $5.2 \pm 0.7$ & 16.8 & $5.82 \pm 0.09$ & $2.09 \pm 0.18$ & $2.06 \pm 0.18$ & 90 & 156 & RF \\
\enddata
\tablecomments{Columns are as follows: (1) Galaxy designation; (2) Hubble morphological type ($T$-type) from HyperLeda \citep{HyperLeda}; (3) Adopted distance \citep{Wiegert15}; (4) Stellar mass \citep{Li16}; (5) Global star formation rate derived from H$\alpha$ and WISE 22 $\mu$m luminosities (R.~Walterbos et al. 2026, in prep.); (6) SFR surface density, calculated using the global SFR and the 22 $\mu$m physical diameter (R.~Walterbos et al. 2026, in prep.); (7) Inclination angle \citep{Irwin12, Krause18}; (8) Position angle of the major axis \citep{Krause18, HyperLeda}; (9) AGN classification based on \citet{Li16}: RF (radio faint), RB (radio bright; unresolved radio core with polarization or inverted spectrum), and ERB (extremely radio bright; showing additional evidence of jets, bubbles, or flux variability).}
\end{deluxetable*}

\subsection{Thermal Emission Subtraction and Nonthermal Spectral Index}\label{subsec:thermal_sub}

Radio continuum emission comprises thermal free-free and nonthermal synchrotron components. To isolate the nonthermal emission and calculate the nonthermal spectral index ($\alpha_{\rm nth}$), we estimated the thermal contribution for each valid pixel from the SFR maps. We utilized the empirical relation from \citet{Vargas19}, adopting a typical electron temperature of $T_e = 10^4\,{\rm K}$. The resulting thermal map was subtracted from each total-intensity map to produce nonthermal maps at all four frequencies. This thermal conversion assumes optically thin free-free emission. These estimates also inherit systematic uncertainties from the hybrid SFR maps, for which we apply no additional correction for possible mid-infrared contributions from older or evolved stellar populations. The effects of finite free-free opacity and SFR-tracer uncertainties on the thermal subtraction and resolved correlations are discussed in Section~\ref{subsec:systematics}.

Based on the nonthermal maps at 144~MHz, 1.575~GHz, 3.0~GHz, and 6.0~GHz, we performed a pixel-by-pixel least-squares fit to derive the spatially resolved nonthermal spectral index, defined by the convention $I_{\nu} \propto \nu^{\alpha_{\rm nth}}$. For each pixel, we recorded the fitting errors and the coefficient of determination ($r^2$) of this power-law fit. To ensure the reliability of the spectral index maps, pixels with a poor fit ($r^2 < 0.9$) were systematically excluded from further analysis. In practice, this criterion rejected fewer than 4\% of valid pixels in the worst case (NGC~3079) and zero pixels in 17 of the 19 galaxies, confirming that the four-band power-law model is an adequate description for the vast majority of the data.

\subsection{SFR Surface Densities}\label{subsec:sfr_sd}

We defined the projected SFR surface density for each pixel as $\Sigma_{\rm SFR} = {\rm SFR}_{\rm pixel} / A_{\rm pixel}$, where $A_{\rm pixel}$ is the physical area of a pixel. For edge-on galaxies, each pixel integrates emission along the full line-of-sight depth through the galaxy. Since this effective emitting depth is not directly constrained by the observations, our derived $\Sigma_{\rm SFR}$ is strictly a projected surface density, uncorrected for the geometric path length.

To investigate the spatially resolved radio--SFR relation, we compared the hybrid SFR surface density ($\Sigma_{\rm SFR}^{\rm hyb}$, i.e.\ the $\Sigma_{\rm SFR}$ defined above, with the superscript retained here only to distinguish it from the radio-derived value) with the radio-derived SFR surface density ($\Sigma_{\rm SFR}^{\rm RC}$). The latter was converted from the nonthermal intensity $I_{\nu}$ using the modified version of the \citet{Condon92} relation introduced by \citet{Heesen24}, which assumes a constant radio spectral index of $\alpha = -0.8$:
\begin{equation}
    \Sigma_{\rm SFR}^{\rm RC} = \left(\frac{\nu}{1.4\,{\rm GHz}}\right)^{0.8} \left(\frac{\theta_{\rm FWHM}}{\rm arcsec}\right)^{-2} \left(\frac{3310\,I_{\nu}}{\rm Jy\,beam^{-1}}\right),
\end{equation}
where $\theta_{\rm FWHM}$ is the beam FWHM in arcseconds.

We retain this radio-based SFR convention to facilitate comparison with \citet{Heesen24} and to express the radio measurements on a common SFR scale across the sample. Here, $\Sigma_{\rm SFR}^{\rm RC}$ denotes the nonthermal intensity converted to SFR units, without implying that all emission within a resolution element traces local, in situ star formation.

\subsection{Fitting Procedure}\label{subsec:fitting}

We investigate two spatially resolved scaling relations. The first is the dependence of the nonthermal spectral index on the SFR surface density. The nonthermal spectral index is sensitive to CRE aging during propagation away from injection sites in star-forming regions. Previous studies have investigated this connection on global and spatially resolved scales using semi-logarithmic relations between the nonthermal spectral index and star-formation tracers \citep{Tabatabaei17,Tabatabaei22}. Here, we adopt such an empirical relation to provide a first-order description of the trend over the sampled range of SFR surface density:
\begin{equation}\label{eq:alpha_sfr}
    \alpha_{\rm nth} = a\,\log_{10}\Sigma_{\rm SFR} + b.
\end{equation}
The second is the radio--SFR correlation,
\begin{equation}\label{eq:rc_sfr}
    \log_{10}\Sigma_{\rm SFR}^{\rm RC} = k\,\log_{10}\Sigma_{\rm SFR}^{\rm hyb} + m,
\end{equation}
with both surface densities in units of $10^{-3}\,M_{\odot}\,{\rm yr}^{-1}\,{\rm kpc}^{-2}$. Equation~(\ref{eq:rc_sfr}) is fitted independently for each of the four frequency bands.

Both relations are fitted for each galaxy individually using Orthogonal Distance Regression (ODR) in \texttt{SciPy}, which accounts for measurement uncertainties on both axes. We additionally compute the Spearman rank correlation coefficient $r_s$ and the reduced chi-square $\chi^2_\nu$ as goodness-of-fit diagnostics.

To derive representative properties for the entire sample, we employ bootstrap resampling to estimate the medians and the $16^{\rm th}$/$84^{\rm th}$ percentiles of the fitted parameters and statistical metrics. The individual galaxy fits for the $\alpha_{\rm nth}$--$\Sigma_{\rm SFR}$ relation are presented in Figure~\ref{fig:alpha_sfr_grid}, while the ensemble results are detailed in Table~\ref{tab:alpha_sfr_fits} and visualized in Figure~\ref{fig:alpha_sfr_fits}. Analogously, the individual galaxy fits for the spatially resolved radio--SFR correlation across the four frequency bands are listed in Table~\ref{tab:rc_hyb_fits_pergal}, while the median fitting results are summarized in Table~\ref{tab:rc_hyb_fits_median} and Figure~\ref{fig:sfrsd_rc_vs_hyb}.

\section{Results}\label{sec:result}

\subsection{The $\alpha_{\rm nth}$--$\Sigma_{\rm SFR}$ Relation}\label{subsec:alpha_sfr}

\begin{figure*}[ht!]
    \centering
    \includegraphics[width=\linewidth]{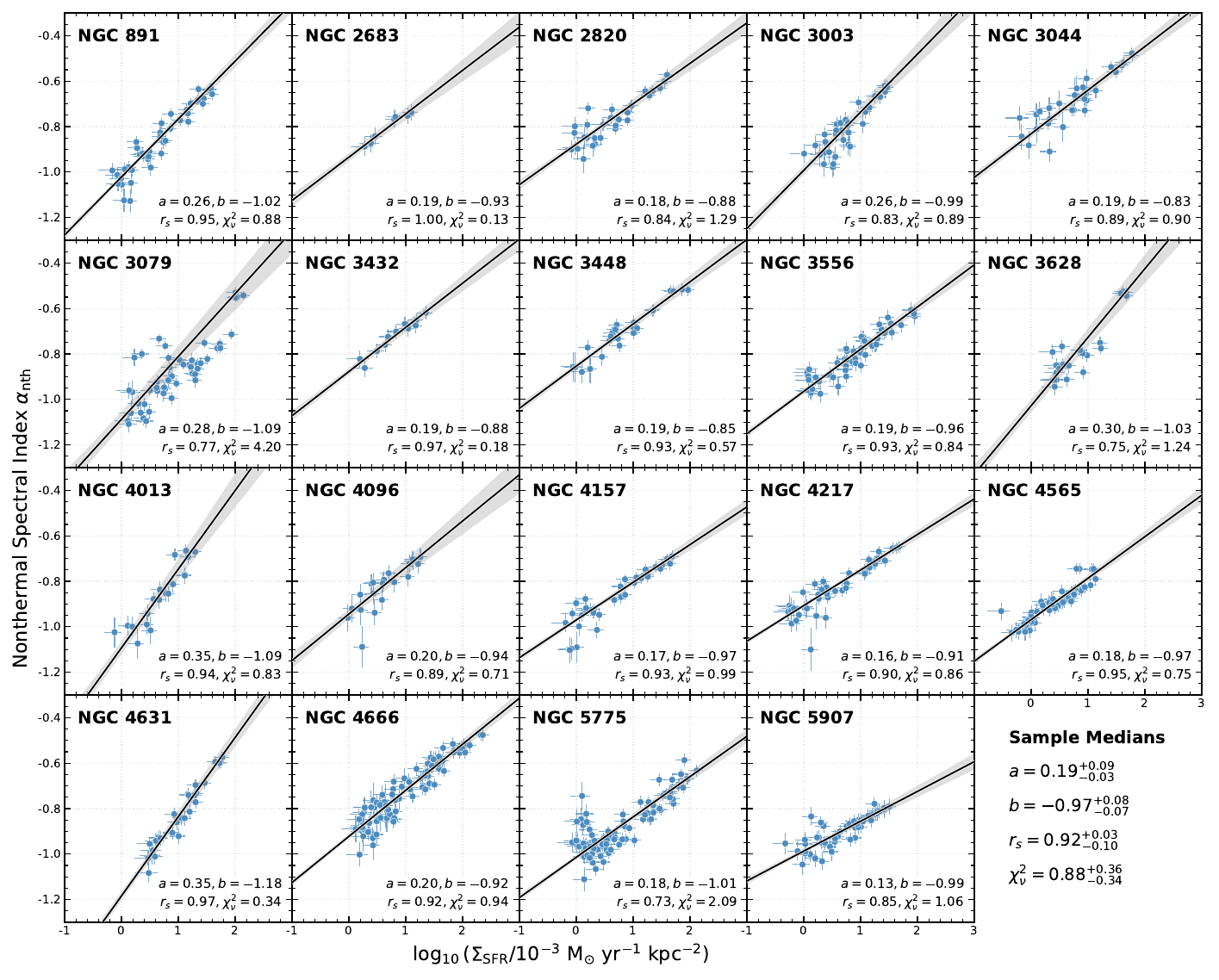}
    \caption{Spatially resolved relation between $\alpha_{\rm nth}$ and $\Sigma_{\rm SFR}$ across the galaxy sample. In each panel, points show the pixel-by-pixel measurements for an individual galaxy, with error bars indicating the $1\sigma$ uncertainties on both axes. Solid lines and gray shaded regions denote the individual best-fitting ODR models and their corresponding $1\sigma$ intervals. The bottom-right panel summarizes the median relation for the entire sample.}
    \label{fig:alpha_sfr_grid}
\end{figure*}

\begin{figure}[ht!]
    \centering
    \includegraphics[width=\linewidth]{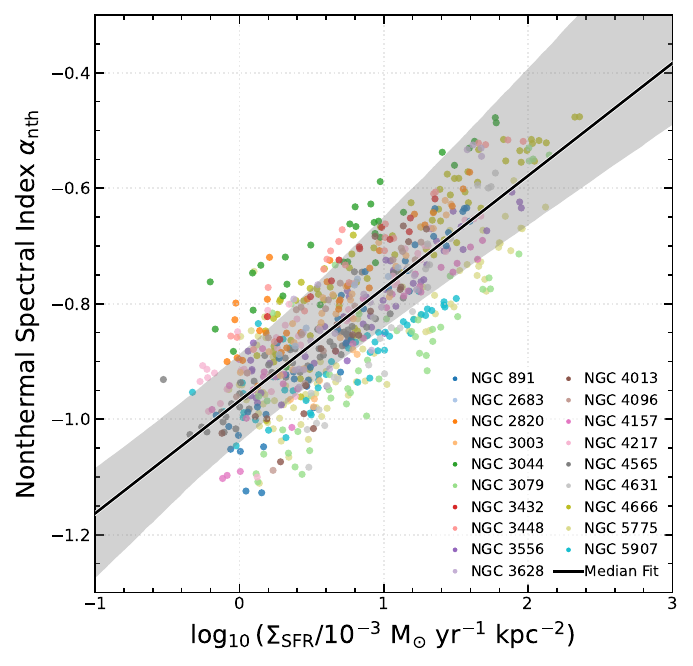}
    \caption{Spatially resolved relation between $\alpha_{\rm nth}$ and $\Sigma_{\rm SFR}$. Data points are color-coded by galaxy. The sample median ODR fit is shown as a solid black line, with the gray band indicating the $16^{\rm th}$--$84^{\rm th}$ bootstrap percentile interval. Details of the best-fitting relations are summarized in Table \ref{tab:alpha_sfr_fits}.}
    \label{fig:alpha_sfr_fits}
\end{figure}

Star-forming regions exhibit relatively flat synchrotron spectra, consistent with the recent injection of young CREs into the surrounding ISM \citep[e.g.,][]{Tabatabaei13}. As these CREs propagate away from their supernova birthplaces and lose energy through synchrotron and IC cooling, the synchrotron spectrum steepens progressively, as observed toward interarm regions and the outer disk \citep{Tabatabaei07}, and with increasing height above the disk in edge-on galaxies \citep[e.g.,][]{Schmidt19,Stein19b,Heald22,Stein23,Xu25}.
The nonthermal spectral index $\alpha_{\rm nth}$ therefore serves as a direct diagnostic of where, within a galaxy, CRE injection dominates over cooling. For edge-on systems, the high inclination enables us to trace the cooling progress from the star-forming midplane to the radio halo. In this work, we use the spatially resolved $\alpha_{\rm nth}$--$\Sigma_{\rm SFR}$ relation as a diagnostic tool to identify regions of active CRE injection and to analyse the disk-to-halo transport mechanism. Figures~\ref{fig:alpha_sfr_grid} and \ref{fig:alpha_sfr_fits} show the pixel-by-pixel relation between the nonthermal spectral index $\alpha_{\rm nth}$ and the hybrid SFR surface density $\Sigma_{\rm SFR}$ for all 19 galaxies at the common physical resolution of ${\sim}2.1\,{\rm kpc}$.

The nonthermal synchrotron spectrum becomes flatter with increasing $\Sigma_{\rm SFR}$ in all 19 galaxies (Table~\ref{tab:alpha_sfr_fits}). Under our adopted convention, the Spearman rank correlation coefficients span $r_s = 0.73$--$1.00$, with a sample median of $r_s = 0.92_{-0.10}^{+0.03}$, and all $p$-values are $< 0.01$. The reduced chi-square values yield a sample median of $\chi^2_\nu = 0.88_{-0.34}^{+0.36}$, indicating that the ODR linear model provides an adequate description of the data in most cases. The lowest value occurs for NGC~2683 ($\chi^2_\nu = 0.13$); as both the closest and physically smallest galaxy, it is sampled by the fewest independent pixels ($N_{\rm pix} = 7$), so its fit is highly sensitive to individual points and should be interpreted with caution. The highest value occurs for NGC~3079 ($\chi^2_\nu = 4.20$), for which the elevated residuals are attributable to the presence of an active galactic nucleus (AGN), associated nuclear radio bubbles, and a biconical outflow \citep{Li19,Li24,Hodges-Kluck20}. These features contribute excess synchrotron emission in the central region that is not accounted for by the SFR-based model, systematically displacing those pixels from the best-fit relation.

The ODR fitting yields individual slopes spanning $a = 0.13$--$0.35$, with a sample median of $a = 0.19_{-0.03}^{+0.09}$ (Table~\ref{tab:alpha_sfr_fits}). This indicates a general trend where the synchrotron spectrum is systematically flatter in active star-forming regions. This is physically expected, as regions of elevated $\Sigma_{\rm SFR}$ are sites of active supernova-driven CRE injection, where the freshly accelerated electron population maintains a relatively flat nonthermal spectrum ($\alpha_{\rm nth} \sim -0.5$ to $-0.6$). Crucially, since these dense star-forming regions also host amplified magnetic fields that theoretically accelerate synchrotron cooling ($t_{\rm sync} \propto B^{-3/2}$), the persistence of flat spectra unambiguously demonstrates that \textit{in-situ} young CRE injection overwhelmingly dominates over local cooling losses. Conversely, regions of low $\Sigma_{\rm SFR}$, which predominantly correspond to the extraplanar halo, are dominated by aged CREs that have undergone significant synchrotron and IC cooling, producing a steeper spectrum ($\alpha_{\rm nth} \sim -0.9$ to $-1.1$). Interestingly, our median slope of ${\sim}0.19$ is highly consistent with both the global relation reported for nearby galaxies \citep[$0.17\pm0.06$;][]{Tabatabaei17} and the spatially resolved relation found in the star-forming regions of the face-on galaxy M~33 \citep[$0.21\pm0.01$;][]{Tabatabaei22}. Despite the severe line-of-sight integration inherent to edge-on systems, which inevitably mixes the emission from active star-forming spiral arms (flat spectra) and diffuse inter-arm regions (steep spectra) across the galactic disk, the intrinsic local correlation between CRE injection and SFR surface density remains remarkably robust.

Despite the overall consistency of the sample, the scatter in the fitted slope substantially exceeds the individual fitting uncertainties, ranging from $a = 0.13$ (NGC~5907) to $a = 0.35$ (NGC~4631). To determine whether this scatter is driven by global galaxy properties, we performed Spearman rank correlation tests between the fitted slopes and several global parameters: $M_*$, SFR, ${\rm SFR}/M_{*}$, $\Sigma_{\rm SFR}$, and morphological $T$-type. None of these parameters shows a statistically significant correlation with the slope at the $p < 0.05$ level, and all yield $|r_s| < 0.25$. The absence of a strong dependence on any single global parameter implies that the slope of the $\alpha_{\rm nth}$--$\Sigma_{\rm SFR}$ relation reflects a local, within-galaxy contrast between injection and cooling regions. This gradient is likely governed by a complex interplay of dynamical environments, nuclear activity, and CRE transport mechanisms (see Section~\ref{subsec:cr_transport} for a detailed discussion). 

For instance, NGC~4631, which exhibits the steepest slope in the sample ($a = 0.35 \pm 0.02$), is a well-known interacting system embedded in a complex tidal environment involving NGC~4627 and NGC~4656 \citep{Wang23}. It hosts one of the most prominent radio halos among the CHANG-ES galaxies \citep{Mora-Partiarroyo19a,Mora-Partiarroyo19b}, with extraplanar emission extending to $\gtrsim 6\,{\rm kpc}$ above the midplane at $L$-band. The tidal interaction likely enhances the efficiency of vertical CRE transport, amplifying the contrast in $\alpha_{\rm nth}$ between the star-forming disk and the extended halo, thereby producing the steepest $\alpha_{\rm nth}$--$\Sigma_{\rm SFR}$ slope. At the opposite extreme, NGC~5907 ($a = 0.13 \pm 0.01$) is a quiescent galaxy characterized by a deficiency of nonthermal radio emission \citep{Dumke00}. Its radio halo, while present \citep[$S$-band scale height $h_{\rm halo} = 2.02 \pm 0.29\,{\rm kpc}$;][]{Heesen25}, is less prominent relative to its large optical disk \citep[$R_e = 11.4 \pm 0.4\,{\rm kpc}$;][]{Heesen25}, resulting in a low halo-to-disk ratio and consequently a weaker spatial variation in $\alpha_{\rm nth}$ across the galaxy.

We note that NGC~5775 shows enhanced $\alpha_{\rm nth}$ scatter at the low-$\Sigma_{\rm SFR}$ end. These pixels lie at the outermost edge of the valid coverage region. Given that NGC~5775 has the largest halo scale height \citep[$h_{\rm halo} = 3.17\pm0.06$;][]{Heesen25} in our sample and an exceptionally extended multiphase halo \citep[e.g.,][]{Li08,Heald22}, our analysis reaches these faint outer regions in this galaxy. Their large per-pixel uncertainties and symmetric distribution about the disc suggest reduced measurement fidelity rather than genuine $\alpha_{\rm nth}$ structure.

\begin{deluxetable}{lccccc}
\tablecaption{ODR Fitting Results for $\alpha_{\rm nth}$-$\Sigma_{\rm SFR}$ relation\label{tab:alpha_sfr_fits}}
\tablehead{
\colhead{Galaxy} & \colhead{$N_{\rm pix}$} & \colhead{Slope ($a$)} & \colhead{Intercept ($b$)} & \colhead{$r_s$} & \colhead{$\chi^2_\nu$}
}
\colnumbers
\startdata
NGC~891 & $42$ & $0.26 \pm 0.01$ & $-1.02 \pm 0.01$ & $0.95$ & $0.88$ \\
NGC~2683 & $7$ & $0.19 \pm 0.02$ & $-0.93 \pm 0.02$ & $1.00$ & $0.13$ \\
NGC~2820 & $30$ & $0.18 \pm 0.02$ & $-0.88 \pm 0.01$ & $0.84$ & $1.29$ \\
NGC~3003 & $32$ & $0.26 \pm 0.02$ & $-0.99 \pm 0.02$ & $0.83$ & $0.89$ \\
NGC~3044 & $31$ & $0.19 \pm 0.01$ & $-0.83 \pm 0.01$ & $0.89$ & $0.90$ \\
NGC~3079 & $49$ & $0.28 \pm 0.03$ & $-1.09 \pm 0.03$ & $0.77$ & $4.20$ \\
NGC~3432 & $12$ & $0.19 \pm 0.02$ & $-0.88 \pm 0.01$ & $0.97$ & $0.18$ \\
NGC~3448 & $23$ & $0.19 \pm 0.01$ & $-0.85 \pm 0.01$ & $0.93$ & $0.57$ \\
NGC~3556 & $51$ & $0.19 \pm 0.01$ & $-0.96 \pm 0.01$ & $0.93$ & $0.84$ \\
NGC~3628 & $22$ & $0.30 \pm 0.03$ & $-1.03 \pm 0.03$ & $0.75$ & $1.24$ \\
NGC~4013 & $19$ & $0.35 \pm 0.03$ & $-1.09 \pm 0.03$ & $0.94$ & $0.83$ \\
NGC~4096 & $20$ & $0.20 \pm 0.03$ & $-0.94 \pm 0.02$ & $0.89$ & $0.71$ \\
NGC~4157 & $34$ & $0.17 \pm 0.01$ & $-0.97 \pm 0.01$ & $0.93$ & $0.99$ \\
NGC~4217 & $43$ & $0.16 \pm 0.01$ & $-0.91 \pm 0.01$ & $0.90$ & $0.86$ \\
NGC~4565 & $49$ & $0.18 \pm 0.01$ & $-0.97 \pm 0.01$ & $0.95$ & $0.75$ \\
NGC~4631 & $27$ & $0.35 \pm 0.02$ & $-1.18 \pm 0.02$ & $0.97$ & $0.34$ \\
NGC~4666 & $80$ & $0.20 \pm 0.01$ & $-0.92 \pm 0.01$ & $0.92$ & $0.94$ \\
NGC~5775 & $84$ & $0.18 \pm 0.01$ & $-1.01 \pm 0.01$ & $0.73$ & $2.09$ \\
NGC~5907 & $50$ & $0.13 \pm 0.01$ & $-0.99 \pm 0.01$ & $0.85$ & $1.06$ \\
\tableline
\multicolumn{2}{c}{Median} & $0.19_{-0.03}^{+0.09}$ & $-0.97_{-0.07}^{+0.08}$ & $0.92_{-0.10}^{+0.03}$ & $0.88_{-0.34}^{+0.36}$ \\
\enddata
\tablecomments{Columns are as follows: (1) Galaxy designation. (2) Number of independent pixels ($N_{\rm pix}$) within the valid fitting region, each corresponding to one beam-sized resolution element. (3) and (4) Best-fit slope ($a$) and intercept ($b$) derived from the ODR fit, modeled as $\alpha_{\rm nth} = a \log_{10}(\Sigma_{\rm SFR} / 10^{-3}\,M_{\odot}\,{\rm yr}^{-1}\,{\rm kpc}^{-2}) + b$. (5) Spearman's rank correlation coefficient ($r_s$); all corresponding $p$-values are $< 0.01$. (6) Reduced chi-square ($\chi^2_\nu$). The bottom row presents the sample medians, with asymmetric uncertainties representing the $16^{\rm th}$ and $84^{\rm th}$ percentiles evaluated via bootstrap resampling.}
\end{deluxetable}

\subsection{Frequency Dependence of the Radio--SFR Relation}\label{subsec:radio_sfr_freq}

\begin{figure*}[ht!]
    \centering
    \includegraphics[width=\linewidth]{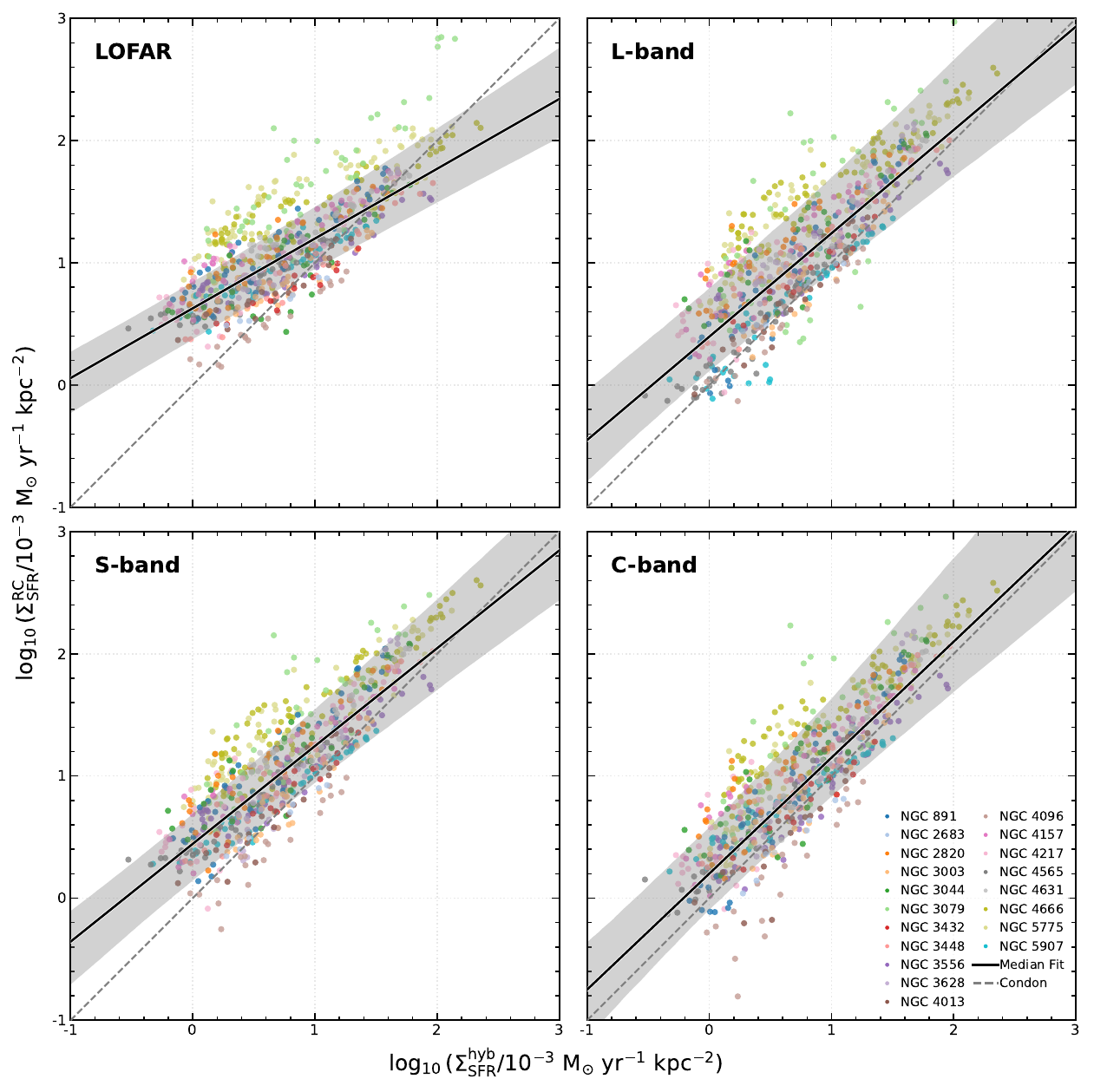}
    \caption{Radio versus hybrid $\Sigma_{\rm SFR}$ scaling relations. The panels display the radio-derived $\Sigma_{\rm SFR}$ at 144~MHz (LOFAR), 1.575~GHz ($L$-band), 3.0~GHz ($S$-band), and 6.0~GHz ($C$-band), respectively, as a function of the hybrid $\Sigma_{\rm SFR}$. Data points are color-coded by galaxy. The dashed gray line indicates the $1:1$ relation from \citet{Condon92}, while the solid black line and the shaded region represent the median fit and its $16^{\rm th}$--$84^{\rm th}$ bootstrap percentile interval, respectively. The best-fitting parameters are listed in Table \ref{tab:rc_hyb_fits_median}.}
    \label{fig:sfrsd_rc_vs_hyb}
\end{figure*}

\begin{figure}[ht!]
    \centering
    \includegraphics[width=\linewidth]{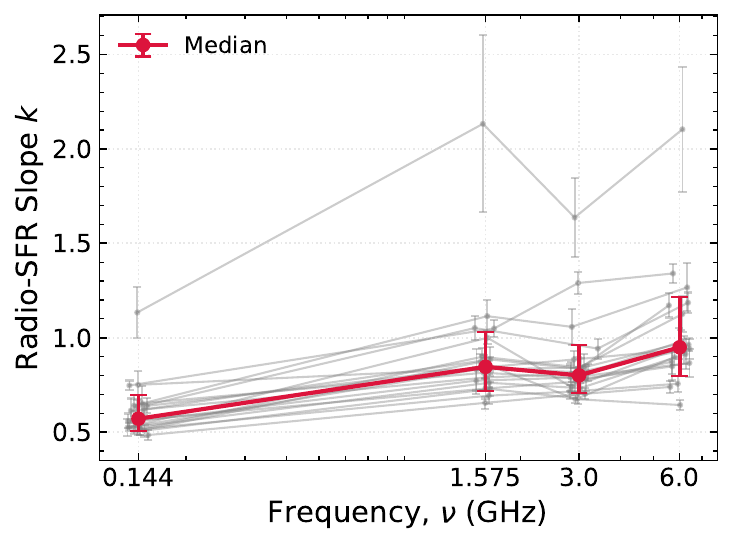}
    \caption{Radio-SFR relation slope $k$ as a function of frequency $\nu$. Individual galaxy fits are shown as thin gray lines, with small horizontal offsets applied to each point for visual clarity. The sample median is indicated by the solid red line, where the error bars represent the $16^{\rm th}$–$84^{\rm th}$ percentile interval of the bootstrap distribution.}
    \label{fig:slope_vs_freq}
\end{figure}

In Figure~\ref{fig:sfrsd_rc_vs_hyb}, we present the spatially resolved relations between the nonthermal radio-derived SFR surface density ($\Sigma_{\rm SFR}^{\rm RC}$) and the hybrid SFR surface density ($\Sigma_{\rm SFR}^{\rm hyb}$) across all four frequency bands. For each galaxy in every band, the ODR fits yield well-constrained power-law relations, with Spearman's rank correlation coefficients ranging from $r_s = 0.72$ to $1.00$ and $p$-values $\ll 0.01$. The best-fit slopes and intercepts for individual galaxies in all four bands are listed in Table~\ref{tab:rc_hyb_fits_pergal}. To characterize the overall population while mitigating the influence of extreme individual cases, we adopt the sample median of the best-fit parameters. The most notable outlier is NGC~3079, which exhibits anomalously steep slopes across all bands (e.g., $k = 1.13 \pm 0.14$ at 144~MHz) and consistently elevated reduced chi-square values (e.g., $\chi^2_\nu = 3.48$ at 144~MHz). As discussed in Section~\ref{subsec:alpha_sfr}, this galaxy hosts a known AGN with associated bubble-driven outflows that enhance the radio luminosity well beyond what is expected from star formation alone \citep{Li19}. Apart from this outlier, the individual slopes are distributed about the sample median rather than pulled down by a few flat galaxies (e.g., $14$ of the $19$ galaxies lie between $0.74$ and $1.19$ at $C$-band).

Figure~\ref{fig:slope_vs_freq} reveals a clear overall steepening of the median slope $k$ as a function of observing frequency, rising from $0.57_{-0.07}^{+0.13}$ in the LOFAR 144~MHz band to $0.95_{-0.15}^{+0.26}$ in the $C$-band (6~GHz). Although there is an apparent slight decrease from the $L$-band ($0.85_{-0.13}^{+0.19}$) to the $S$-band ($0.80_{-0.09}^{+0.16}$), these values are statistically indistinguishable within their $16^{\rm th}$--$84^{\rm th}$ bootstrap percentile intervals. This minor fluctuation likely reflects residual systematic differences between the two datasets, arising from variations in $uv$-coverage and imaging pipelines, rather than a genuine physical reversal of the underlying trend.

To assess the influence of radio-bright nuclear activity, we recalculated the sample medians from the existing per-galaxy fits after excluding the four RB/ERB systems identified by \citet{Li16}: NGC~3079, NGC~3628, NGC~4013, and NGC~4565. The median slopes become $0.55$, $0.81$, $0.78$, and $0.91$ at 144~MHz, $L$-, $S$-, and $C$-band, respectively, differing from the full-sample values by less than $0.05$ in all bands. The overall steepening with frequency is therefore preserved without these systems.

The physical origin of this frequency gradient lies in the energy-dependent cooling of CREs. The synchrotron cooling timescale is strongly dependent on the electron energy ($E$), scaling as $t_{\rm cool} \propto E^{-1} \propto \nu^{-1/2}$ \citep{Condon92}. High-frequency radio continuum ($C$-band) primarily traces highly energetic CREs with shorter radiative lifetimes. These electrons radiate away their energy rapidly before they can diffuse significantly from their injection sites, and thus remain tightly confined to active star-forming regions. Consequently, high-frequency radio maps closely track the spatial distribution of recent CRE injection, and the spatially resolved radio--SFR relation is correspondingly steeper ($k \sim 1$). Even this near-linear slope remains below the super-linear exponent of the globally integrated relation, which additionally reflects the dependence of the synchrotron emissivity on magnetic field strength \citep{Niklas97}; we return to this comparison in Section~\ref{subsec:viewing_angle}. In contrast, low-frequency emission (e.g., LOFAR 144~MHz) traces an older, lower-energy CRE population with much longer cooling timescales. These electrons can diffuse extensively over kiloparsec scales into the extended radio halo \citep[e.g.,][]{Krause18, Mora-Partiarroyo19a, Stein19, Stein20, Stein23, Xu25, Heesen25}. The resulting spatial smearing of the radio emission effectively decouples the localized radio continuum from the current local SFR, flattening the resolved relation to $k \approx 0.57$ at 144~MHz. This observed frequency gradient is therefore consistent with energy-dependent CRE cooling and transport mechanisms within the ISM.

When compared to the spatially resolved LOFAR study by \citet{Heesen24}, which predominantly features face-on and moderately inclined systems, our median slopes show a similar frequency-dependent trend but are systematically higher (their sample median slopes are roughly $0.50$ at 144~MHz and $0.63$ at 1.36~GHz). A crucial distinction is that our sample consists entirely of edge-on galaxies. In an edge-on viewing geometry, the line-of-sight integration inevitably blends the compact, actively star-forming regions within the midplane with the diffuse, older emission extending into the inter-arm regions and the lower halo. This geometric projection forces our spatially resolved measurements to partially mimic ``global'' integration characteristics, which as noted above are super-linear, thereby steepening the slope. We will explore this geometric projection effect in detail in Section~\ref{subsec:viewing_angle}.

\begin{deluxetable}{lcccc}
\tablecaption{Median Fitting Results for Radio-SFR relation\label{tab:rc_hyb_fits_median}}
\tablehead{
\colhead{Band} & \colhead{Slope ($k$)} & \colhead{Intercept ($m$)} & \colhead{$r_s$} & \colhead{$\chi^2_\nu$}
}
\colnumbers
\startdata
LOFAR & $0.57_{-0.07}^{+0.13}$ & $0.63_{-0.25}^{+0.22}$ & $0.92_{-0.08}^{+0.04}$ & $1.34_{-0.82}^{+1.37}$ \\
$L$-band & $0.85_{-0.13}^{+0.19}$ & $0.40_{-0.29}^{+0.39}$ & $0.94_{-0.06}^{+0.02}$ & $1.28_{-0.82}^{+1.18}$ \\
$S$-band & $0.80_{-0.09}^{+0.16}$ & $0.44_{-0.32}^{+0.25}$ & $0.93_{-0.05}^{+0.03}$ & $1.14_{-0.68}^{+1.05}$ \\
$C$-band & $0.95_{-0.15}^{+0.26}$ & $0.20_{-0.30}^{+0.38}$ & $0.94_{-0.05}^{+0.02}$ & $0.92_{-0.51}^{+1.21}$ \\
\enddata
\tablecomments{Columns are as follows: (1) Frequency Band. (2) and (3) Median ODR best-fit slope ($k$) and intercept ($m$), modeled as $\log_{10}\Sigma_{\rm SFR}^{\rm RC} = k \log_{10}\Sigma_{\rm SFR}^{\rm hyb} + m$, with both surface densities in units of $10^{-3}\,M_{\odot}\,{\rm yr}^{-1}\,{\rm kpc}^{-2}$. (4) Median Spearman's rank correlation coefficient ($r_s$). (5) Median reduced chi-square ($\chi^2_\nu$). The asymmetric uncertainties represent the $16^{\rm th}$ and $84^{\rm th}$ percentiles derived via bootstrap resampling.}
\end{deluxetable}

\section{Discussion}\label{sec:discussion}

\subsection{Disk-to-Halo CRE Transport}\label{subsec:cr_transport}

The systematic flattening of the nonthermal synchrotron spectrum with increasing $\Sigma_{\rm SFR}$, detected across all 19 galaxies (Section~\ref{subsec:alpha_sfr}), provides direct, spatially resolved evidence for CRE energy losses during propagation from the star-forming disk into the radio halo. Regions of high $\Sigma_{\rm SFR}$, primarily concentrated along the galactic midplane, are sites of active supernova-driven CRE injection. Here, freshly accelerated electrons have not yet cooled significantly, yielding a characteristically flat nonthermal spectrum. As CREs escape into the extended halo, where $\Sigma_{\rm SFR}$ is low, synchrotron radiation and IC scattering progressively steepen the local spectral index, producing the steep, diffuse synchrotron spectrum ($\alpha \sim -0.5$ to $-1.1$) characteristic of galaxy halos \citep{Li16}. The observed $\alpha_{\rm nth}$--$\Sigma_{\rm SFR}$ relation, with a sample median slope of $a = 0.19_{-0.03}^{+0.09}$, therefore directly traces this contrast between injection and cooling within individual galaxies.

This spatially resolved perspective naturally explains why many previous studies relying on galaxy-integrated properties failed to detect a significant correlation. For instance, \citet{Li16} found no significant trend between the integrated $C$- to $L$-band spectral index and global SFR in the CHANG-ES sample, attributing the tight scatter in $\alpha$ to closely coupled emission at these two frequencies. \citet{Heesen22} similarly reported no correlation between the integrated 144~MHz--1.4~GHz spectral index and $\Sigma_{\rm SFR}$. The absence of a detectable signal in both cases reflects the fact that global flux integration blends flat-spectrum midplane emission with steep-spectrum diffuse halo emission, erasing the spatial gradients that our pixel-by-pixel analysis resolves. Even structurally similar samples yield only a moderate global correlation \citep{Smolinski26}. Notably, when thermal contamination is explicitly removed via broad-band ($1$--$10$~GHz) spectral energy distribution fitting, \citet{Tabatabaei17} recovered a global nonthermal slope of $0.17 \pm 0.06$, in close agreement with our spatially resolved median. This consistency suggests that the intrinsic link between nonthermal emission and recent star formation is preserved across scales once thermal contamination is properly accounted for, and that spatially resolved analysis is required to unmask it.

The edge-on geometry of our sample introduces a qualitative shift in the physical interpretation of this correlation relative to face-on studies. In the face-on galaxy M~33, \citet{Tabatabaei22} measured a resolved nonthermal slope of $\sim$0.21--0.39, primarily reflecting horizontal diffusion and the short residency times of CREs within individual star-forming complexes. Consequently, the observed $\alpha_{\rm nth}$--$\Sigma_{\rm SFR}$ slope provides insight into the projected emission variations associated with the vertical escape and transport of CREs. The edge-on viewing angle changes the two-dimensional projection of the underlying three-dimensional emission distribution, making extraplanar emission more apparent relative to midplane emission. Our measured slopes therefore offer clues to baryonic feedback and cosmic-ray-driven winds that extend perpendicular to the galactic plane \citep{Heesen18}.

The broad scatter in individual fitted slopes ($a = 0.13$--$0.35$), which shows no dependence on global host-galaxy parameters (Section~\ref{subsec:alpha_sfr}), points instead to the local dynamical environment as the primary modulator of vertical CRE transport efficiency. In addition to the local dynamical environment, the large-scale organization of the vertical magnetic field likely plays a complementary role in regulating CRE escape efficiency. Both factors are most clearly illustrated by the extremes of our sample.

Among the steepest slopes, three galaxies share a common thread of tidal interaction. NGC~4631 ($a = 0.35 \pm 0.02$) is a violently interacting system embedded in a complex tidal environment involving NGC~4627 and NGC~4656 \citep{Wang23}, and hosts one of the most prominent radio halos in the CHANG-ES sample, with extraplanar emission extending $\gtrsim 6$~kpc at $L$-band \citep{Mora-Partiarroyo19b}. NGC~3628 ($a = 0.30 \pm 0.03$), a member of the Leo Triplet with a 140~kpc tidal tail \citep{Haynes79}, hosts an advection-dominated radio halo \citep{Heesen18}. For these two galaxies, tidal perturbations likely enhance the efficiency of vertical outflows, amplifying the spectral contrast between the injection-dominated disk and the cooling-dominated halo. The steep slope of NGC~4631 may additionally reflect the exceptionally well-ordered X-shaped vertical magnetic field revealed by polarimetric observations \citep{Mora-Partiarroyo19a}. Such a coherent field geometry likely traces the vertical outflow that shapes it, providing an independent indication of the vigorous vertical transport in this system. NGC~4013 ($a = 0.35 \pm 0.03$) represents a complementary case: despite its low SFR and a comparatively weak radio halo showing no radio signature of ongoing interaction \citep{Stein19b}, its pronounced warp and tidal stellar stream extending $\sim 3$~kpc above the midplane \citep{Martinez09} supply the extraplanar structure that steepens the resolved $\alpha_{\rm nth}$--$\Sigma_{\rm SFR}$ relation, reinforcing that the slope tracks the local dynamical structure rather than the global properties of the galaxy. At the opposite extreme, NGC~5907 ($a = 0.13 \pm 0.01$) is a dynamically quiescent isolated galaxy lacking evidence for starburst-driven winds or coherent vertical magnetic field structure, with a comparatively weak radio halo relative to its large stellar disk \citep{Dumke00}, producing a correspondingly shallow spatial variation in $\alpha_{\rm nth}$.

The presence of nuclear activity introduces an additional layer of complexity in the resolved CRE transport. Following the classification scheme of \citet{Li16}, our sample contains four radio-bright (RB) or extremely radio-bright (ERB) systems: NGC~3079, NGC~3628, NGC~4013, and NGC~4565. For NGC~3079 and NGC~3628, the $\alpha_{\rm nth}$--$\Sigma_{\rm SFR}$ distributions depart from a single clean relation: the bulk of the pixels follow a shallow baseline trend broadly consistent with the sample median, while a few points at the very highest $\Sigma_{\rm SFR}$, corresponding to the nuclear regions, lie well above it with anomalously bright radio emission. These radio-excess points reflect the superposition of two physically distinct components: a baseline component governing standard disk-to-halo cosmic-ray transport, typical of normal star-forming galaxies, and a highly active nuclear region dominated by massive outflows. These central outflows---whether driven by AGN bubbles \citep[in NGC~3079;][]{Li19} or a starburst nucleus \citep[in NGC~3628;][]{Fabbiano80}---inject complex, highly modified CRE populations that deviate from the standard cooling trend. We note that, although the bubble/outflow emission is physically offset from the star-forming midplane, at our $2.1\,{\rm kpc}$ pixel scale the two are blended within the central resolution elements and cannot be spatially separated. In contrast, no such radio-excess nuclear points are observed in the RB galaxies NGC~4013 and NGC~4565. This absence implies that their radio-bright cores are currently less active in driving massive outflows; indeed, CRE transport in both galaxies is reported to be dominated by slow advection or diffusion with velocities $\lesssim 100\,{\rm km\,s^{-1}}$ \citep{Stein19b,Xu26}. In these systems, transport mechanisms remain more uniform across the galaxy, preventing the appearance of such nuclear outliers. Excluding these four RB/ERB systems changes the sample median slope of the $\alpha_{\rm nth}$--$\Sigma_{\rm SFR}$ relation only from $0.195$ to $0.189$; excluding NGC~3079 alone yields $0.193$. Thus, although nuclear emission may affect individual galaxy fits, the sample median slope remains stable when these systems are excluded.

In summary, the spatially resolved $\alpha_{\rm nth}$--$\Sigma_{\rm SFR}$ relation offers a powerful diagnostic tool for tracing the transport and cooling of CREs in edge-on galaxies. While global relations often suffer from line-of-sight blending, the edge-on view makes the projected disk-to-halo spectral gradient more apparent. The baseline slope represents the standard transport efficiency of typical galaxies, whereas substantial deviations (scatter or slope breaks) serve as sensitive indicators of environmental disruptions, such as tidal interactions or centralized outflow activities. Furthermore, this spatially resolved physical picture provides a natural context for evaluating the inclination-dependent biases in low-frequency SFR calibrations, which we explore in Section~\ref{subsec:viewing_angle}.

\subsection{Inclination Effects on Radio-based SFR Calibrations}\label{subsec:viewing_angle}

Our spatially resolved radio--SFR slopes are systematically steeper than those reported for face-on galaxies at comparable physical resolutions. By combining all pixels across a morphologically diverse sample (inclinations $7\degr$--$77\degr$) at a 1.2~kpc resolution, \citet{Heesen24} reported joint-fit slopes of $0.86 \pm 0.02$ at 144~MHz and $0.92 \pm 0.01$ at 1.4~GHz. However, such combined fits are heavily weighted by the most pixel-rich galaxies and may not reflect typical intra-galaxy variations. To establish a more meaningful baseline, we recalculated the per-galaxy median slopes from their individual fits, yielding $\sim 0.50$ at 144~MHz and $\sim 0.63$ at 1.4~GHz. While our edge-on median at 144~MHz ($0.57_{-0.07}^{+0.13}$) is broadly consistent with their face-on counterpart, our $L$-band slope ($0.85_{-0.13}^{+0.19}$) is notably steeper. This widening discrepancy at higher frequencies points to a strong geometric effect inherent to the edge-on perspective.

This geometric steepening is driven by severe line-of-sight blending. In edge-on systems, any line of sight passing through the star-forming midplane simultaneously intersects a substantial column of the extended radio halo. This vertical integration physically mixes the flat-spectrum, localized emission of young CREs with the steep-spectrum, diffuse halo emission that has long decoupled from recent star formation \citep[e.g.,][]{Murphy08, Heesen18, Krause18}. Furthermore, azimuthal projection forces a single synthesized beam to co-add active spiral arms and quiescent inter-arm regions along the disk depth. Together, these projections effectively erase the distinct sub-kpc morphological contrast that spatially resolved face-on studies exploit.

Consequently, edge-on resolved slopes are shifted away from purely local (face-on) values and toward global, galaxy-integrated scaling relations. Global radio--SFR slopes are consistently near-linear to super-linear. At GHz frequencies, \citet{Heesen14} reported a nonthermal slope of $1.16 \pm 0.08$, and \citet{Li16} derived $1.13 \pm 0.07$ at 1.6~GHz and $1.06 \pm 0.07$ at 6~GHz for the CHANG-ES edge-on sample, while the nonthermal slopes of \citet{Tabatabaei17} correspond to $k \sim 1.1$ at both 20~cm and 6~cm. At 144--150~MHz, the reported values span a wider range, from $1.02 \pm 0.07$ \citep{Heesen24}, $1.058 \pm 0.007$ \citep{Smith21}, and $1.07 \pm 0.01$ \citep{Gurkan18} to $1.402 \pm 0.072$ \citep{Heesen22}. Such super-linearity is expected if the synchrotron emissivity depends on the magnetic field as well as on the CRE injection rate, as in the equipartition model of \citet{Niklas97}, which predicts an exponent of $1.3 \pm 0.3$. Our resolved edge-on medians (e.g., $0.85$ at $L$-band; $0.95$ at $C$-band) systematically occupy the intermediate space between the face-on resolved regime ($k \sim 0.6$) and the global regime ($k \gtrsim 1.1$). The line-of-sight integration effectively ``globalises'' the resolved measurement without fully reproducing the integrated flux of the entire galaxy. This local-to-global transition has recently been quantitatively reconciled by \citet{Heesen26}, who showed that global and local radio--SFR measurements can be unified within a single relation once the radio spectral index is incorporated as a second parameter.

This projection framework naturally explains why the face-on versus edge-on discrepancy is much larger at 1.4~GHz than at 144~MHz. At low frequencies, CRE diffusion lengths are so extensive that the radio emission forms a smooth, extended halo regardless of viewing angle; thus, the resolved slope is universally flattened ($\sim 0.5$) in both edge-on and face-on galaxies. At higher frequencies, however, CREs cool faster and trace the structural contrast of the disk much more tightly. A face-on view successfully resolves this arm/inter-arm contrast (yielding $k \approx 0.63$), whereas an edge-on view azimuthally integrates through these distinct structures, collapsing the contrast and driving the slope toward the global near-linear relation ($k \approx 0.85$).

The sub-unity slopes found in resolved studies carry direct consequences for utilizing the radio continuum as a quantitative SFR tracer. A slope $k < 1$ dictates that the radio emission systematically overestimates $\Sigma_{\rm SFR}$ in faint, quiescent regions---where the beam is heavily contaminated by diffuse halo or inter-arm electrons---while underestimating it in the brightest star-forming complexes. This compression effect is most severe at 144~MHz ($k \approx 0.57$) and becomes progressively milder at higher frequencies.

We conclude that $C$-band (and by extension, $\sim 5$--$10$~GHz) observations provide the most reliable spatially resolved SFR calibrations for edge-on or highly inclined systems, as their slopes naturally approach linearity ($k \approx 0.95$) within the sample scatter. Conversely, low-frequency resolved calibrations require explicit corrections for inclination-dependent halo contamination before they can be safely applied as linear SFR tracers.

\subsection{Dependence of the Resolved Slope on Global Star Formation Activity}\label{subsec:calorimetry}

The resolved radio--SFR slopes listed in Table~\ref{tab:rc_hyb_fits_pergal} vary considerably from galaxy to galaxy, and it is natural to ask whether this variation is set by the overall star formation activity of the host. The balance between CRE escape and synchrotron losses may depend on the global $\Sigma_{\rm SFR}$. For advective escape, the escape time is set by the ratio of the radio continuum scale height $h$ to the outflow speed $v$. Since $h$ scales with the radius of the star-forming disk, $h \propto r_\star$ \citep{Krause18}, and the outflow speed rises with star formation activity as $v \propto \Sigma_{\rm SFR}^{0.4}$ \citep{Heesen18,Heesen21}, this gives $t_{\rm esc} \propto r_\star\,\Sigma_{\rm SFR}^{-0.4}$. The synchrotron lifetime scales as $t_{\rm syn} \propto \nu^{-0.5}B^{-1.5} \propto \nu^{-0.5}\Sigma_{\rm SFR}^{-0.5}$, adopting $B \propto \Sigma_{\rm SFR}^{1/3}$, so that $t_{\rm esc}/t_{\rm syn} \propto \nu^{0.5}\,r_\star\,\Sigma_{\rm SFR}^{0.1}$ \citep{Heesen22}. At a given frequency and disk size, this weak increase in $t_{\rm esc}/t_{\rm syn}$ with $\Sigma_{\rm SFR}$ may lead to a slight steepening of the resolved radio--SFR slope. The stronger frequency dependence of $t_{\rm esc}/t_{\rm syn}$ is qualitatively consistent with the steepening discussed in Section~\ref{subsec:radio_sfr_freq}.

We find no statistically significant trend. Spearman rank correlations between the per-galaxy slopes and the global $\Sigma_{\rm SFR}$ of Table~\ref{tab:sample_properties} are insignificant in all four bands ($|r_s| \leq 0.36$, $p > 0.1$). Dividing the sample at the median global $\Sigma_{\rm SFR}$, the low-$\Sigma_{\rm SFR}$ half has slightly steeper median slopes than the high-$\Sigma_{\rm SFR}$ half in the three GHz bands, by factors of $1.18$, $1.04$ and $1.12$ at $L$-, $S$- and $C$-band, while the two halves are indistinguishable at 144~MHz ($0.96 \pm 0.07$). However, these differences are not statistically significant, with a Mann-Whitney test giving $p > 0.09$ in every band. With 19 galaxies and a predicted dependence as weak as $\Sigma_{\rm SFR}^{0.1}$, these results cannot be distinguished from no trend at all, and a larger sample covering a wider range in $\Sigma_{\rm SFR}$ would be needed to test the prediction.

\subsection{Robustness Against Thermal Contamination and Beam Smearing}\label{subsec:systematics}

The analyses presented in Sections~\ref{subsec:alpha_sfr} and \ref{subsec:radio_sfr_freq} are based on nonthermal radio maps convolved to a common physical resolution of $2.1\,{\rm kpc}$. In this section, we evaluate the robustness of our derived scaling relations against two potential sources of systematic uncertainty: thermal continuum subtraction and the adopted physical resolution.

Thermal free-free emission has $\alpha_{\rm th}\approx-0.1$ when optically thin and approaches $\alpha_{\rm th}=+2$ for a homogeneous optically thick source \citep{CondonRansom16}. If left unsubtracted, this contribution artificially flattens the observed total spectral index relative to the true nonthermal value, biasing the $\alpha_{\rm nth}$--$\Sigma_{\rm SFR}$ slope away from zero and pushing the radio--SFR slopes toward linearity. This effect scales with the thermal fraction and is therefore most pronounced at $C$-band, where free-free emission contributes most significantly to the total flux density. To quantify its impact, we repeated all fits using total (thermally unsubtracted) maps at $2.1\,{\rm kpc}$ resolution and compared the results with our baseline nonthermal analysis.

For the $\alpha_{\rm nth}$--$\Sigma_{\rm SFR}$ relation, the sample-median slope shifts from $0.19^{+0.09}_{-0.03}$ (nonthermal) to $0.21^{+0.07}_{-0.03}$ (total), a difference of only $\Delta a = 0.02$, well within the $16^{\rm th}$--$84^{\rm th}$ bootstrap percentile interval of either measurement. The median intercept rises from $-0.97$ to $-0.93$, corresponding to a total spectrum that is slightly flatter than the purely nonthermal spectrum, as expected from residual thermal emission. For the radio--SFR relation, the frequency gradient is fully preserved. The median slopes exhibit only negligible changes across all four bands when comparing the nonthermal to the total emission: LOFAR ($0.57 \to 0.58$), $L$-band ($0.85 \to 0.85$), $S$-band ($0.80 \to 0.82$), and $C$-band ($0.95 \to 0.95$). The median intercepts show a modest overall rise that is largest at $C$-band ($0.20 \to 0.31$), reflecting its higher thermal contribution. These results show that the adopted thermal subtraction does not substantially change the observed scaling relations.

We note that this comparison does not address possible biases in the hybrid SFR tracer itself. Dust emission at $22\,\mu{\rm m}$ heated by older stellar populations may lead to overestimated SFRs \citep{Leroy12}. Because the same SFR maps are used to estimate the thermal component, this contamination could also cause excessive thermal subtraction and bias the inferred nonthermal spectrum toward steeper values. Spatial variations could affect the slopes and scatter of the resolved relations. The data-release study includes non-star-forming infrared emission among the sources of the adopted $\sim30\%$ systematic uncertainty in the hybrid SFR maps, which propagates into the thermal estimates (R.~Walterbos et al.\ 2026, in preparation). Taken together, the tests above indicate that our main results are robust against the adopted thermal-subtraction procedure, while remaining subject to systematic uncertainties in the SFR tracer.

A separate concern is free-free absorption (FFA), which can suppress low-frequency synchrotron emission even after thermal subtraction. The comparison between total and nonthermal maps does not constrain this effect, since both contain the same absorbed synchrotron component. To obtain a conservative estimate of its effect along the extended sightlines through edge-on disks, we use the 22~$\mu$m disk diameters reported by \citet{Wiegert15}. For our 19 galaxies, these span 9.4--36.1~kpc, with a median of 23.4~kpc, motivating a representative full-disk path length of $L=25$~kpc. We adopt an rms electron density of $n_{\rm rms}=0.3\,{\rm cm^{-3}}$ and $T_e=5000$~K to explore an enhanced-absorption case for diffuse ionized gas. The density exceeds representative diffuse warm-ionized-medium values \citep{Haffner09}, while the lower temperature increases the opacity relative to the $10^4$~K adopted for our thermal conversion. These assumptions provide an illustrative enhanced-absorption case for a disk sightline rather than an upper limit for compact nuclear regions. These parameters give ${\rm EM}=n_{\rm rms}^{2}L=2250\,{\rm pc\,cm^{-6}}$ and $\tau_{144}\simeq0.11$. For a fully covering foreground screen, with all synchrotron emission behind the absorbing gas, the attenuation factor is $e^{-\tau_{144}}\simeq0.90$, corresponding to approximately 10\% suppression. If the suppression increases smoothly from negligible at low $\Sigma_{\rm SFR}$ to this value at high $\Sigma_{\rm SFR}$, the approximate slope change over a representative 2~dex range is $\Delta k\simeq\log_{10}(e^{-\tau_{144}})/2\simeq-0.024$, much smaller than the observed difference of 0.38 between 144~MHz and 6~GHz. The corresponding change across the full 144~MHz--6~GHz spectral-index baseline is approximately $\Delta\alpha\simeq-\ln(e^{-\tau_{144}})/\ln(6/0.144)\simeq0.030$. If this change develops over the same 2~dex range in $\Sigma_{\rm SFR}$, its contribution to the fitted $\alpha_{\rm nth}$--$\Sigma_{\rm SFR}$ slope is only $\simeq0.015$, compared with the observed four-band value of $0.19$.

Previous studies also show that absorption signatures depend on the spatial scale considered. The analysis of \citet{Gajovic25} further shows that local low-frequency turnovers in M~101 become less apparent when the emission is averaged over larger regions. This suggests that spatial averaging may also reduce the prominence of localized absorption signatures at our 2.1~kpc resolution. For the estimate above, finite optical depth changes the optically thin thermal prediction by only $1-(1-e^{-\tau_{144}})/\tau_{144}\simeq5\%$. The resulting subtraction error is further reduced by the small thermal contribution at 144~MHz, whereas at higher frequencies the opacity rapidly decreases as $\tau_\nu\propto\nu^{-2.1}$. Although stronger absorption may occur in individual compact regions, these estimates suggest that FFA is unlikely to dominate the observed frequency dependence of the resolved radio--SFR relation.

We next investigate the impact of spatial resolution by degrading our maps to $2.8\,{\rm kpc}$. This larger beam size corresponds to the physical resolution of the LOFAR low-resolution ($20\arcsec$) maps at the distance of NGC~5775, the most distant galaxy in our sample. Smoothing to a larger physical beam blends the star-forming midplane (dominated by fresh CRE injection) with the extended radio halo (dominated by aged, cooling CREs). By mixing these distinct environments, local regions are integrated over larger transport lengths, driving the system locally toward a more global state.

Comparing nonthermal results at $2.1\,{\rm kpc}$ and $2.8\,{\rm kpc}$, we first note that degrading the resolution reduces $N_{\rm pix}$ per galaxy by roughly 30--40\%; NGC~2683 and NGC~3432 are thereby reduced to only four independent pixels and are excluded from the $2.8\,{\rm kpc}$ sample statistics. For the remaining galaxies, the median slope of the $\alpha_{\rm nth}$--$\Sigma_{\rm SFR}$ relation is essentially unchanged, from $0.19^{+0.09}_{-0.03}$ to $0.20^{+0.08}_{-0.05}$. The median intercept rises from $-0.97$ to $-0.93$; this shift reflects the averaging of halo pixels into nominally disk-dominated beams, which flattens the mean spectrum without any change in the underlying thermal fraction. For the radio--SFR relation, median slopes increase modestly across all bands (e.g., from $0.57$ to $0.65$ at LOFAR, and from $0.95$ to $1.01$ at $C$-band), while the frequency gradient is fully preserved.

In summary, neither thermal contamination nor beam smearing materially affects the results of this work. The median $\alpha_{\rm nth}$--$\Sigma_{\rm SFR}$ slope of $\approx 0.19$ and the frequency dependence of the radio--SFR slope are stable under the tests performed here.

\subsection{Impact of Spectral Curvature}\label{subsec:spectral_curvature}

\begin{figure}[ht]
    \centering
    \includegraphics[width=\linewidth]{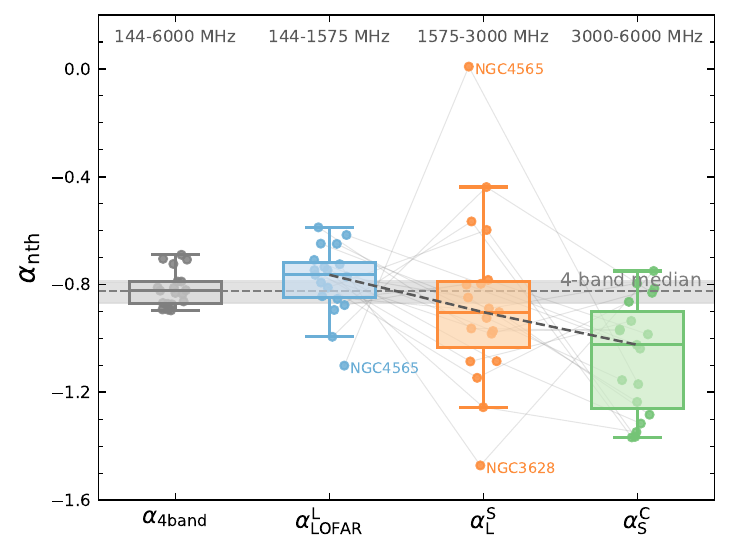}
    \includegraphics[width=\linewidth]{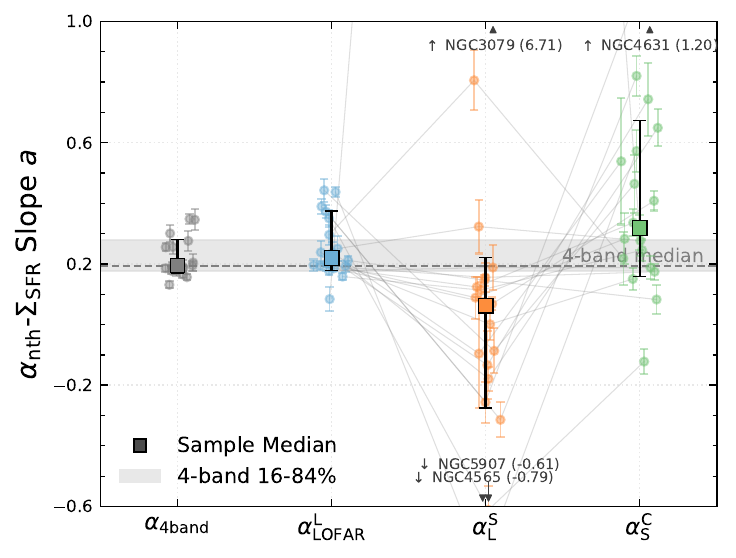}
    \caption{
        Comparison of nonthermal spectral indices ($\alpha_{\rm nth}$) and the fitted slopes ($a$) of the $\alpha_{\rm nth}$--$\Sigma_{\rm SFR}$ relation across different frequency intervals. 
        \textit{Top}: Distributions of the median of $\alpha_{\rm nth}$ derived from the four-band fit ($\alpha_{\rm 4band}$) and adjacent two-point intervals ($\alpha_{\rm LOFAR}^{\rm L}$, $\alpha_{\rm L}^{\rm S}$, $\alpha_{\rm S}^{\rm C}$). Boxes and whiskers represent the interquartile range (IQR) and $1.5\times{\rm IQR}$. Circles mark individual galaxies (jittered for clarity), with faint lines connecting measurements of the same galaxy across intervals. The horizontal dashed line and shaded band indicate the sample median and IQR of $\alpha_{\rm 4band}$.
        \textit{Bottom}: Fitted slopes $a$ evaluated for each spectral index definition. Circles with $1\sigma$ error bars represent individual galaxies. Squares denote the sample medians and their $16^{\rm th}$--$84^{\rm th}$ percentile ranges. The horizontal dashed line and shaded band indicate the median and $16^{\rm th}$--$84^{\rm th}$ percentiles of the four-band slopes, serving as a baseline reference.
    }
    \label{fig:curvature_check}
\end{figure}

The $\alpha_{\rm nth}$--$\Sigma_{\rm SFR}$ relation derived in Section~\ref{subsec:alpha_sfr} relies on a single power-law fit across the full 144~MHz to 6~GHz baseline (hereafter the four-band fit). Over such a wide frequency range, the nonthermal spectrum inherently deviates from a pure power law. Low-frequency emission can be flattened by FFA in the ionized ISM along the line of sight \citep{Israel90}, or by intrinsic spectral curvature shaped by CRE propagation and energy losses \citep{Chyzy18}, while high-frequency emission steepens through synchrotron and IC cooling \citep{Condon92}. Here, we assess whether the resulting spectral curvature introduces a significant bias into our main results.

As shown in the top panel of Figure~\ref{fig:curvature_check}, we compare the four-band results $\alpha_{\rm 4band}$ with two-point spectral indices ($\alpha_{\rm LOFAR}^{\rm L}$, $\alpha_{\rm L}^{\rm S}$, and $\alpha_{\rm S}^{\rm C}$) extracted from adjacent frequency intervals. The sample medians follow a systematic progression: $\alpha_{\rm LOFAR}^{\rm L} = -0.76$, $\alpha_{\rm L}^{\rm S} = -0.90$, and $\alpha_{\rm S}^{\rm C} = -1.02$, with the four-band value $\alpha_{\rm 4band} = -0.82$ falling between the low- and high-frequency extremes. This trend is a direct observational signature of spectral curvature. Given our edge-on viewing geometry, FFA along extended lines of sight through the star-forming disk may contribute to the observed low-frequency flattening. 
Meanwhile, the steepening above 3~GHz is consistent with stronger radiative losses of the aging CREs. The four-band fit therefore represents a physically motivated, frequency-averaged description of the overall CRE energy spectrum.

We test whether this curvature propagates into the $\alpha_{\rm nth}$--$\Sigma_{\rm SFR}$ slope by fitting the relation independently using each two-point index. The results are shown in the bottom panel of Figure~\ref{fig:curvature_check}. The $\alpha_{\rm LOFAR}^{\rm L}$-based fits yield a median slope of $0.22^{+0.15}_{-0.04}$, in agreement with the four-band value of $0.19^{+0.09}_{-0.03}$ within the uncertainties. All 19 galaxies exhibit flatter spectra at higher $\Sigma_{\rm SFR}$, and the trend is statistically significant in 18 cases at $p < 0.01$. This consistency indicates that the inferred spectral flattening is not strongly sensitive to the choice between these two broad frequency baselines.

The $\alpha_{\rm S}^{\rm C}$-based slopes exhibit significant scatter (median $0.32^{+0.36}_{-0.16}$, spanning $-0.12$ to $1.20$). The $\alpha_{\rm L}^{\rm S}$ fits, by contrast, yield far less reliable results: 8 of the 19 galaxies show the opposite spectral trend, and 10 show no statistically significant trend ($p > 0.05$). We attribute this instability to the short frequency separation of the $L$--$S$ and $S$--$C$ intervals (only a factor of $\sim$2 in frequency). Over such narrow frequency intervals, minor flux calibration uncertainties or subtle differences in image resolution are severely amplified, rendering spatially resolved two-point fits unreliable.

In summary, these tests demonstrate the risks of relying on narrow-band pairs to extract resolved spectral indices. The four-band, single power-law approach proves to be a highly stable, noise-resilient estimator. The persistence of the $\sim0.2$ slope across both the four-band and the wide-band LOFAR--$L$ fits assures that our primary finding is robust and physically genuine, regardless of the precise spectral index definition.

\section{Summary and Conclusions}\label{sec:summary}

In this work, we present a spatially resolved, multi-frequency analysis of CRE transport and the local radio--SFR relation in 19 edge-on galaxies, drawing on 144~MHz LOFAR observations from LoTSS-DR3 \citep{LoTSS_DR3}, 1.575, 3.0, and 6.0~GHz VLA observations, and hybrid (H$\alpha$ + 22~$\mu$m) SFR maps from CHANG-ES \citep{Irwin12,Heesen25}. All maps are convolved to a common physical resolution of $2.1\,{\rm kpc}$ to enable a uniform pixel-by-pixel treatment of the nonthermal spectral index $\alpha_{\rm nth}$ and the SFR surface density $\Sigma_{\rm SFR}$. Our goal is to characterize the resolved $\alpha_{\rm nth}$--$\Sigma_{\rm SFR}$ relation as a diagnostic of CRE injection and disk-to-halo transport, and to quantify how the edge-on geometry, combined with this transport, systematically reshapes the frequency-dependent radio--SFR scaling relations that underpin radio-based SFR calibrations. Our main findings are summarized as follows:

\begin{itemize}

\item All 19 galaxies exhibit a statistically significant trend toward flatter nonthermal synchrotron spectra at higher $\Sigma_{\rm SFR}$, with a sample median slope of $a = 0.19_{-0.03}^{+0.09}$. This flattening trend traces the contrast between CRE injection and cooling, with flat spectra marking active midplane injection sites and steeper spectra arising in the extraplanar halo dominated by aged, cooled electrons.

\item The individual slopes of the $\alpha_{\rm nth}$--$\Sigma_{\rm SFR}$ relation span $a = 0.13$ to $0.35$ and show no significant correlation with any global galaxy property. The scatter appears to track the local dynamical environment: tidally perturbed systems (NGC~4631, NGC~3628, NGC~4013) exhibit the steepest slopes, while the quiescent and isolated NGC~5907 shows the shallowest. The most active AGN and starburst hosts further display radio-excess points at the highest $\Sigma_{\rm SFR}$, signalling the superposition of nuclear outflows on the standard disk-to-halo transport.

\item The spatially resolved radio--SFR slope steepens with frequency, rising from $k = 0.57$ at 144~MHz to $k = 0.95$ at 6~GHz. This frequency gradient is consistent with energy-dependent CRE cooling, with high-frequency emission confined near injection sites and low-frequency emission tracing CREs that have diffused into the extended halo.

\item Comparison with the predominantly face-on sample of \citet{Heesen24} reveals that the edge-on geometry imposes a systematic upward shift on the resolved radio--SFR slope, most pronounced at $L$-band. This reflects line-of-sight integration that blends flat-spectrum midplane emission with the steep-spectrum halo and drives the measurement partway toward the global, near-linear regime. The $C$-band (and $\sim 5$--$10$~GHz more broadly) therefore provides the most reliable resolved SFR tracer for highly inclined systems, whereas low-frequency calibrations require explicit corrections for the inclination-dependent dilution of the radio--SFR relation by extended halo emission.

\item The resolved radio--SFR slope shows no significant dependence on the global star formation activity of the host galaxy, although escape-cooling arguments suggest a possible weak trend.

\item The above results are robust against the principal systematic uncertainties of the analysis. Repeating the fits on total (thermally unsubtracted) maps or degrading the resolution to $2.8\,{\rm kpc}$ changes the median $\alpha_{\rm nth}$--$\Sigma_{\rm SFR}$ slope by no more than $\Delta a = 0.02$ and preserves the radio--SFR frequency gradient. The four-band single power-law fit further proves to be a noise-resilient estimator compared to narrow-baseline two-point spectral indices, confirming that our primary findings are not artefacts of unmodeled spectral curvature.

\end{itemize}

Taken together, our results establish the spatially resolved nonthermal spectral index as an observational probe of CRE injection and cooling in edge-on galaxies, and show how CRE transport, combined with the edge-on viewing geometry, is associated with systematic offsets in the frequency-dependent radio--SFR scaling relations. Larger samples spanning a wider range of inclinations, combined with CRE transport modeling, will help to better constrain these inclination-dependent effects in radio SFR calibrations.

\begin{acknowledgments}
J.T.L. acknowledges financial support from the National Natural Science Foundation of China (NSFC) through grants 12321003 and 12273111, from the China Manned Space Program through grants CMS-CSST-2025-A04 and CMS-CSST-2025-A10, and from the Jiangsu Innovation and Entrepreneurship Talent Team Program through grant JSSCTD202436.
J.X. and G.L. acknowledge the research grants from the Ministry of Science and Technology of China (National Key Program for Science and Technology Research and Development, No. 2023YFA1608100), the National Natural Science Foundation of China (No. 12273036), and the research grants from the China Manned Space Project (CMS-CSST-2025-A08).
Y.Y. acknowledges support for this work by the National Natural Science Foundation of China (grant No. 12203098) and the general program of Hunan provincial natural science foundation (grant No. 2026JJ50368)

We acknowledge the use of Gemini \citep{Gemini} for checking grammar and improving the readability of this manuscript.
\end{acknowledgments}

\bibliography{citations}
\bibliographystyle{aasjournalv7}

\appendix

\section{Multi-band Maps and Per-galaxy Fitting Results}
\restartappendixnumbering

\begin{figure*}[ht!]
    \centering
    \includegraphics[width=0.82\linewidth]{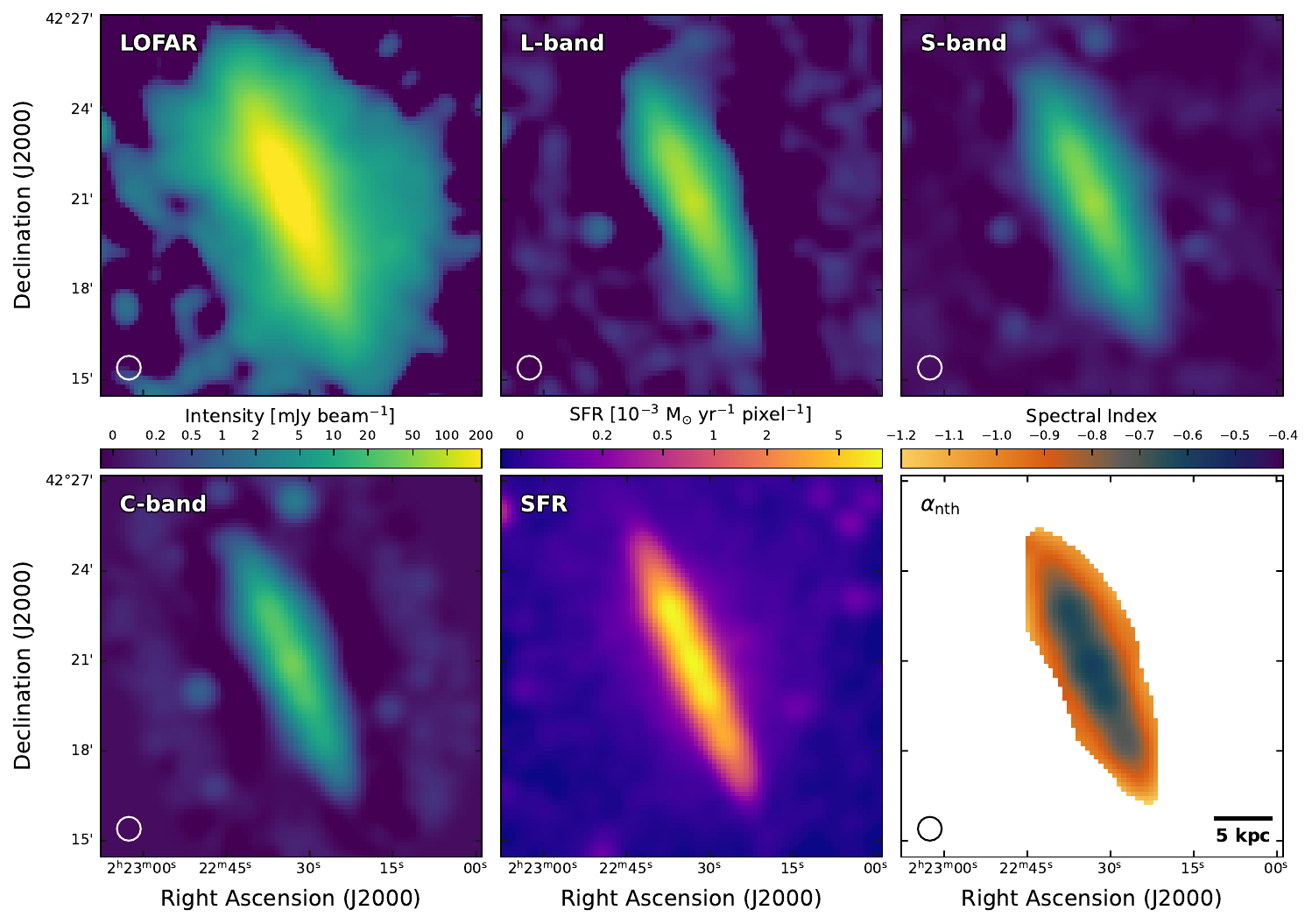}
    \caption{Spatially resolved maps of NGC~891 at a common physical resolution of $2.1\,{\rm kpc}$. The panels show, from left to right and top to bottom: radio continuum emission at 144~MHz (LOFAR), 1.575~GHz ($L$-band), 3.0~GHz ($S$-band), and 6.0~GHz ($C$-band); the hybrid SFR (in units of $10^{-3}\,M_{\odot}\,{\rm yr}^{-1}$); and the nonthermal spectral index ($\alpha_{\rm nth}$). Synthesized beam sizes are indicated in the bottom-left corner of each panel. A $5\,{\rm kpc}$ scale bar is shown in the spectral index panel. The spectral index is visualized with the colormap of \citet{colormap}.}
    \label{fig:exemplar_maps}
\end{figure*}

\begin{figure*}[ht!]
    \centering
    \includegraphics[width=0.82\linewidth]{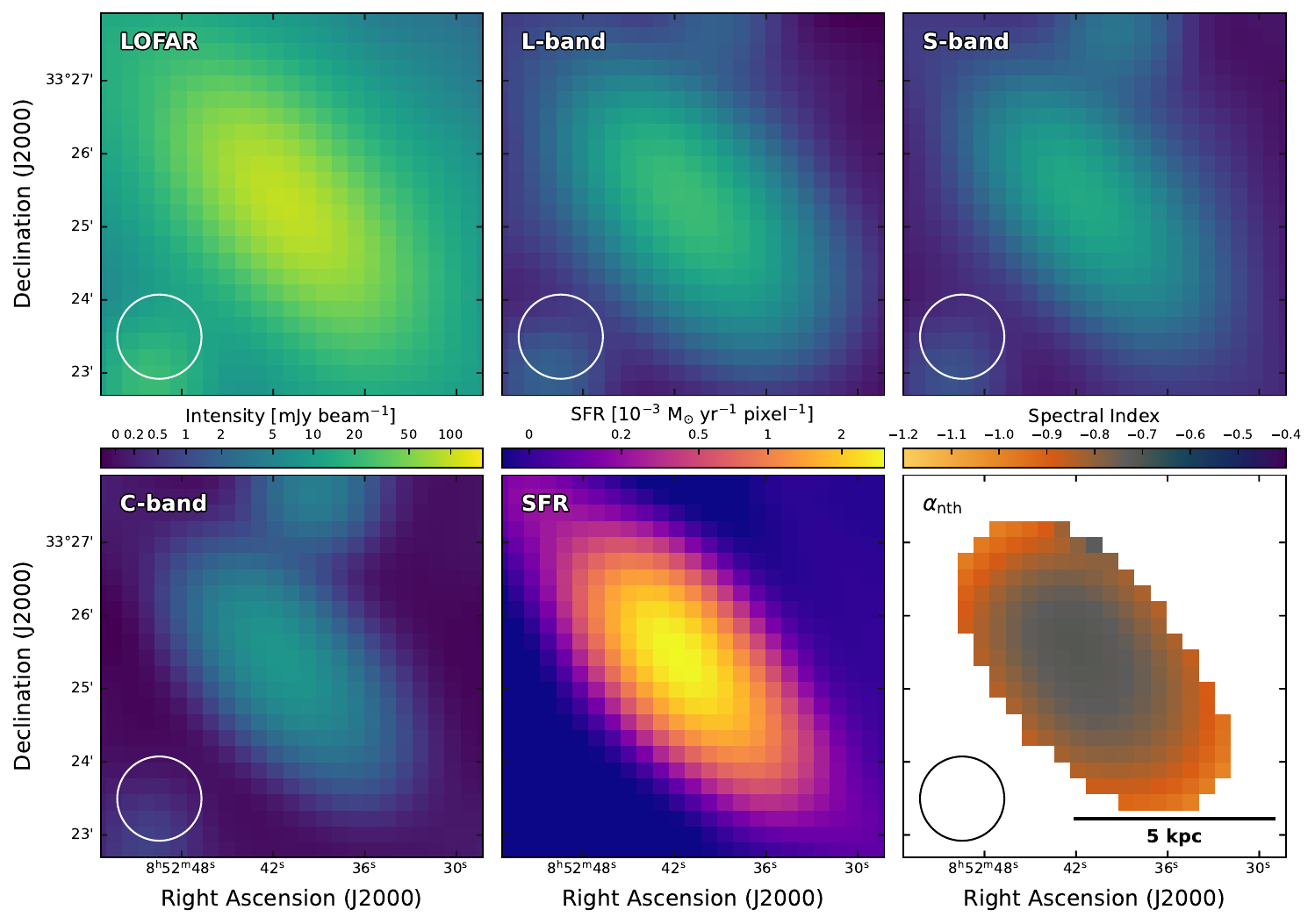}
    \caption{Multi-band maps of NGC~2683. See Figure~\ref{fig:exemplar_maps} for details.}
\end{figure*}

\begin{figure*}[ht!]
    \centering
    \includegraphics[width=0.82\linewidth]{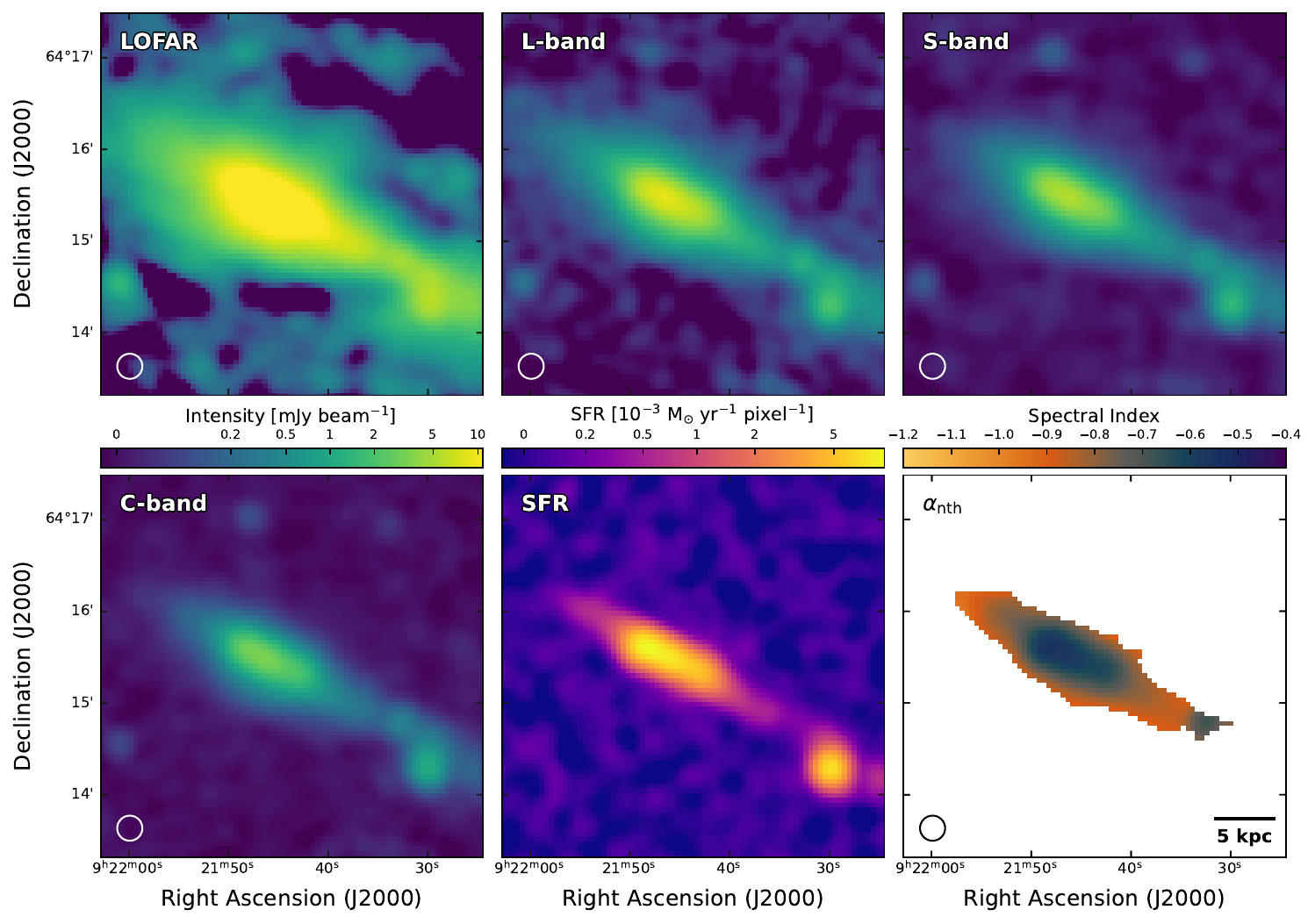}
    \caption{Multi-band maps of NGC~2820. See Figure~\ref{fig:exemplar_maps} for details.}
\end{figure*}

\begin{figure*}[ht!]
    \centering
    \includegraphics[width=0.82\linewidth]{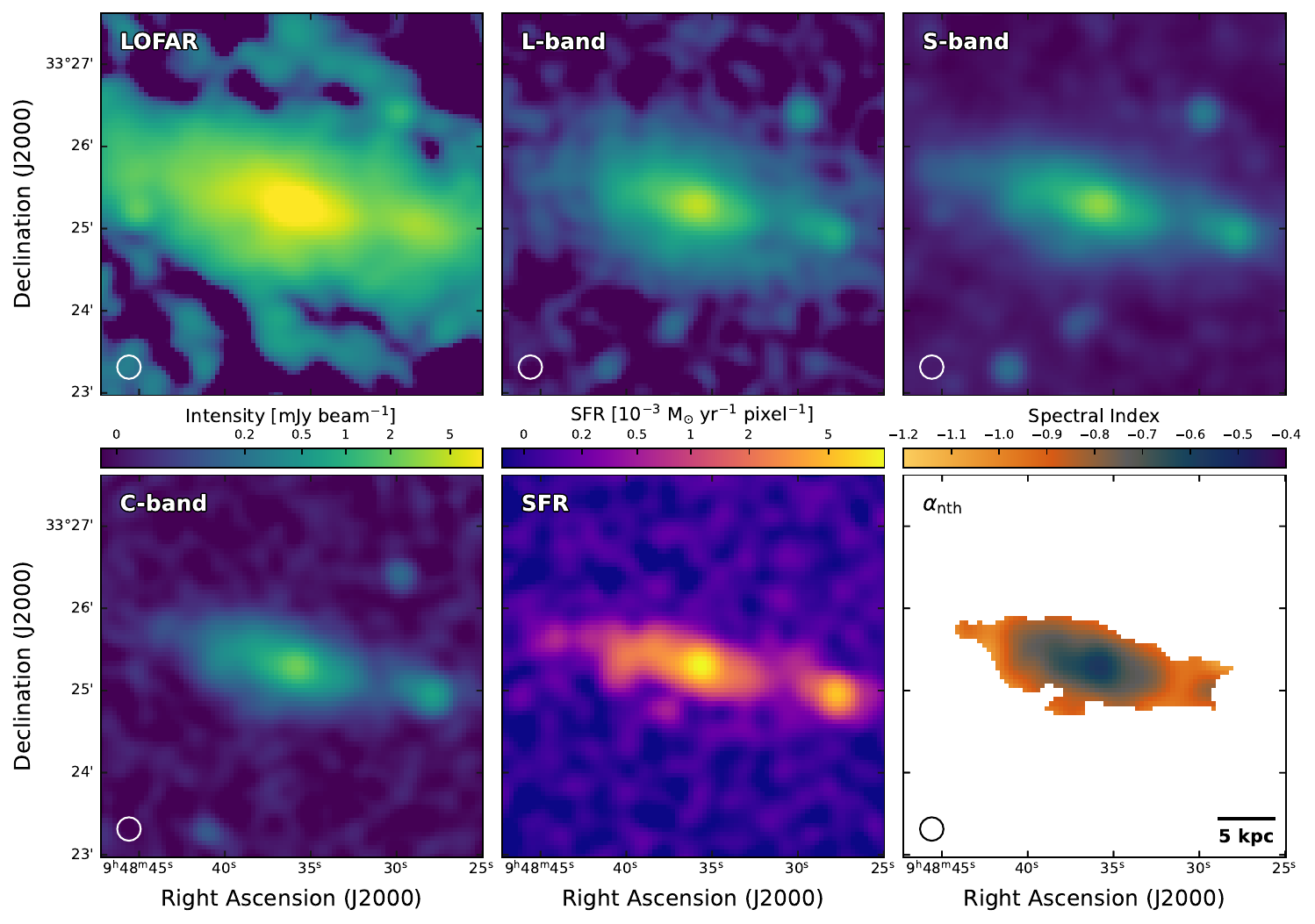}
    \caption{Multi-band maps of NGC~3003. See Figure~\ref{fig:exemplar_maps} for details.}
\end{figure*}

\begin{figure*}[ht!]
    \centering
    \includegraphics[width=0.82\linewidth]{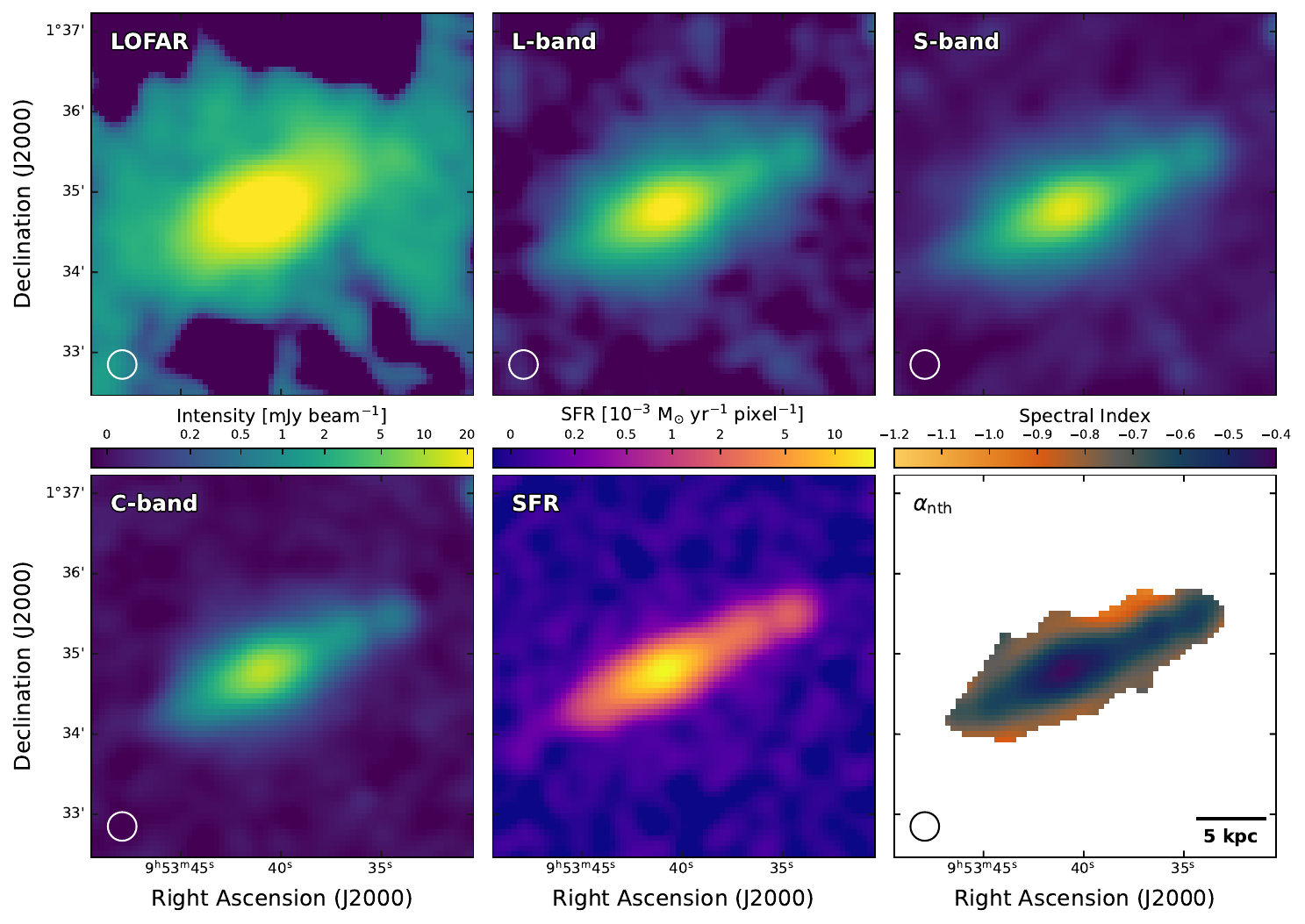}
    \caption{Multi-band maps of NGC~3044. See Figure~\ref{fig:exemplar_maps} for details.}
\end{figure*}

\begin{figure*}[ht!]
    \centering
    \includegraphics[width=0.82\linewidth]{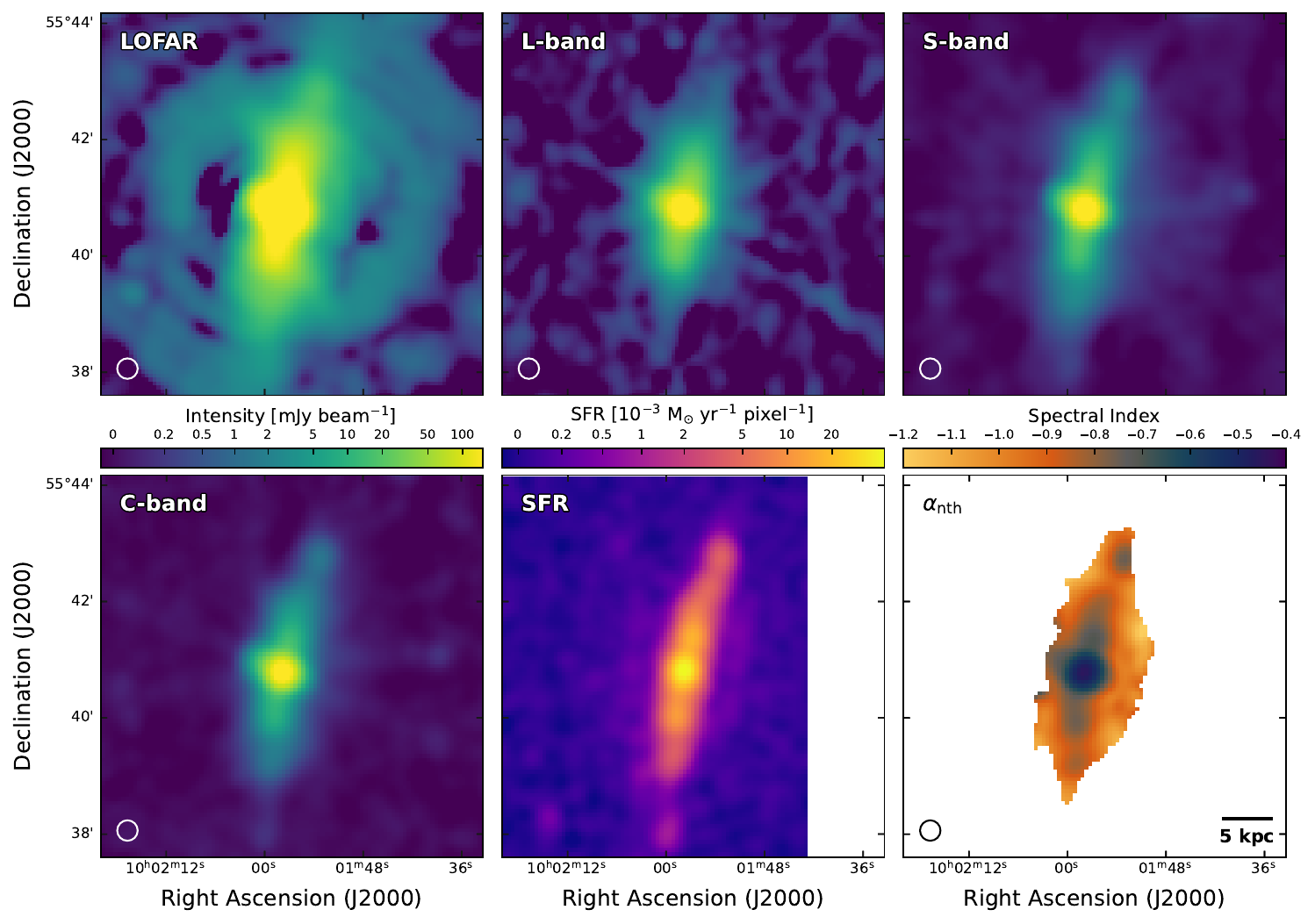}
    \caption{Multi-band maps of NGC~3079. See Figure~\ref{fig:exemplar_maps} for details.}
\end{figure*}

\begin{figure*}[ht!]
    \centering
    \includegraphics[width=0.82\linewidth]{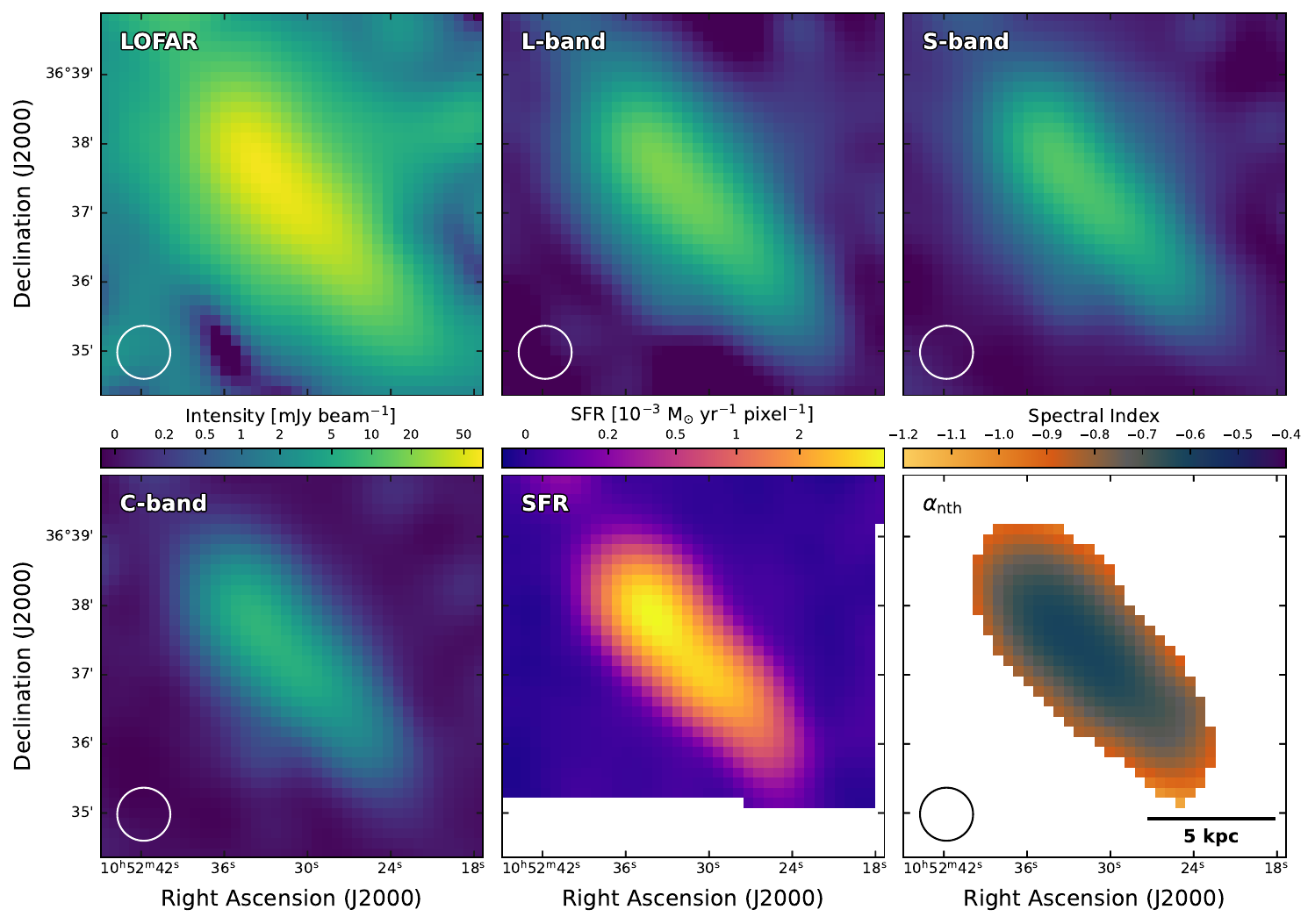}
    \caption{Multi-band maps of NGC~3432. See Figure~\ref{fig:exemplar_maps} for details.}
\end{figure*}

\begin{figure*}[ht!]
    \centering
    \includegraphics[width=0.82\linewidth]{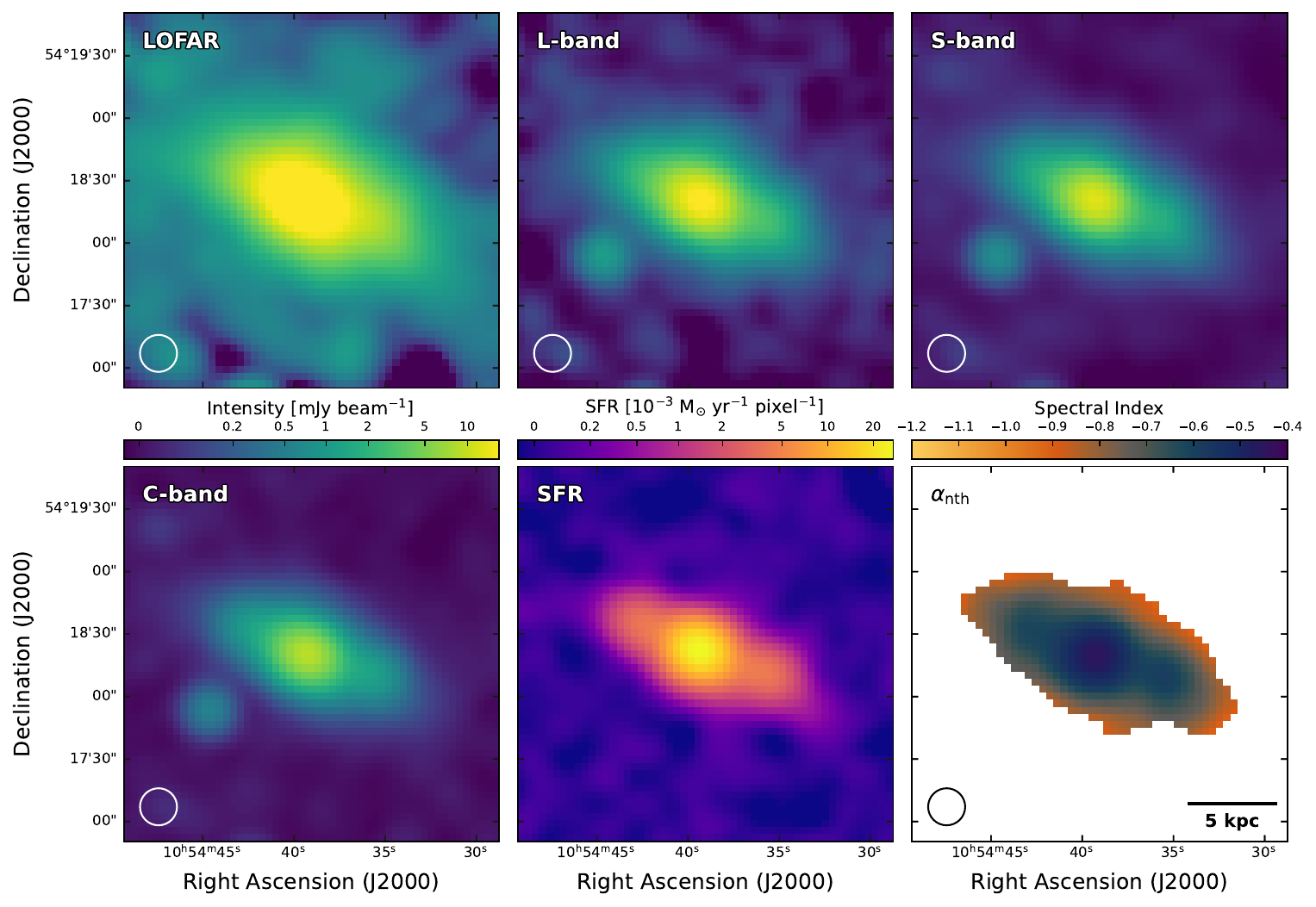}
    \caption{Multi-band maps of NGC~3448. See Figure~\ref{fig:exemplar_maps} for details.}
\end{figure*}

\begin{figure*}[ht!]
    \centering
    \includegraphics[width=0.82\linewidth]{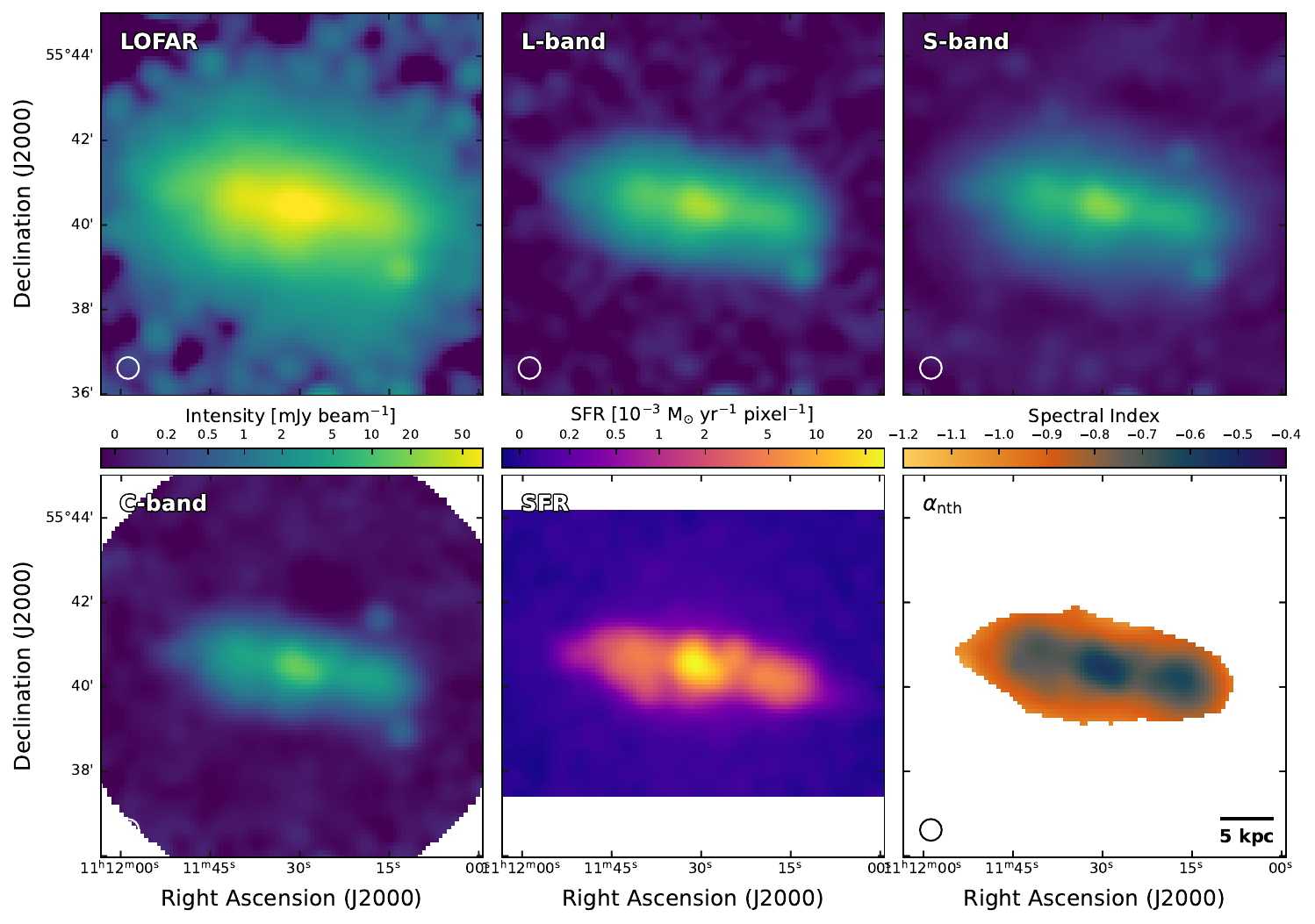}
    \caption{Multi-band maps of NGC~3556. See Figure~\ref{fig:exemplar_maps} for details.}
\end{figure*}

\begin{figure*}[ht!]
    \centering
    \includegraphics[width=0.82\linewidth]{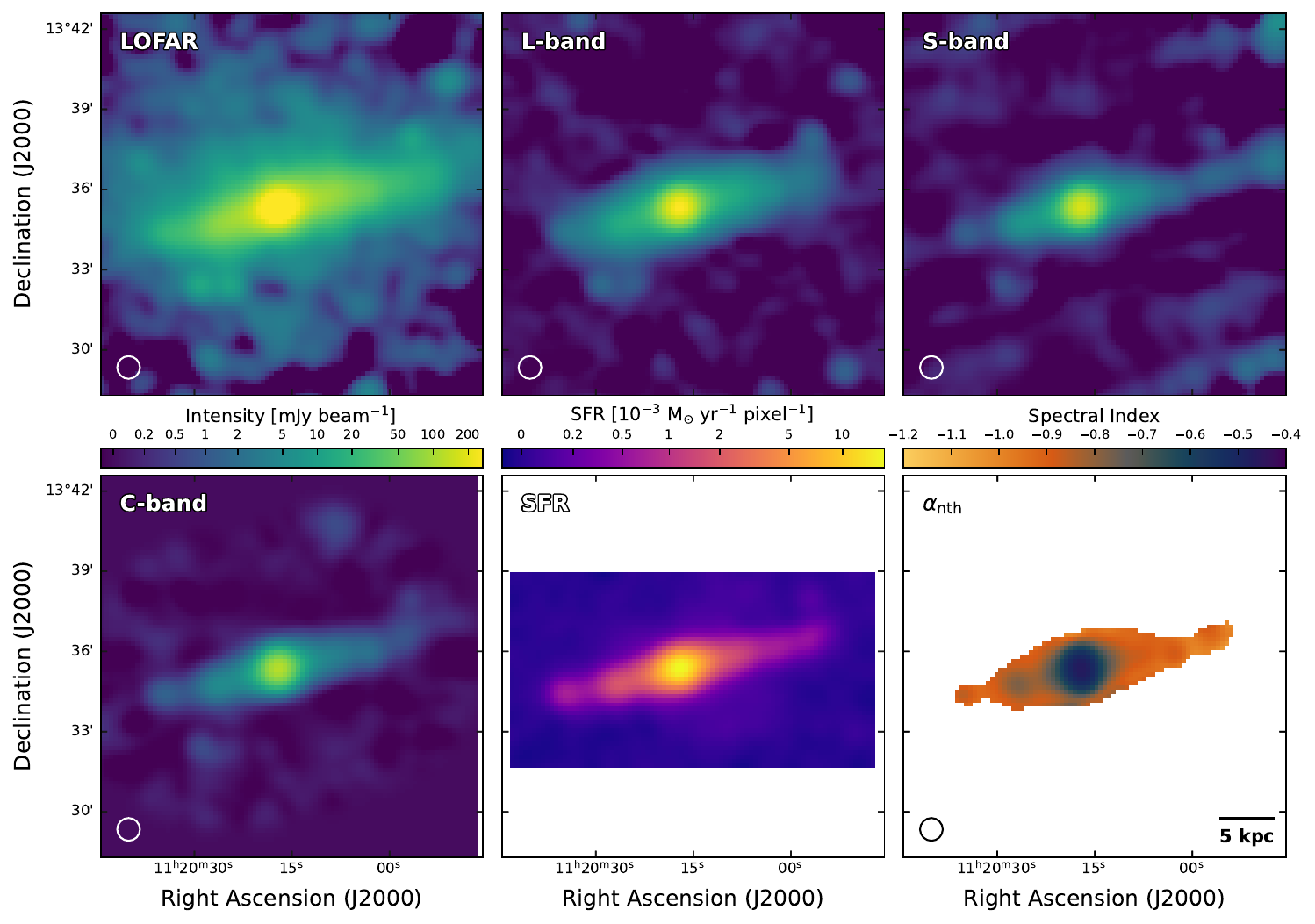}
    \caption{Multi-band maps of NGC~3628. See Figure~\ref{fig:exemplar_maps} for details.}
\end{figure*}

\begin{figure*}[ht!]
    \centering
    \includegraphics[width=0.82\linewidth]{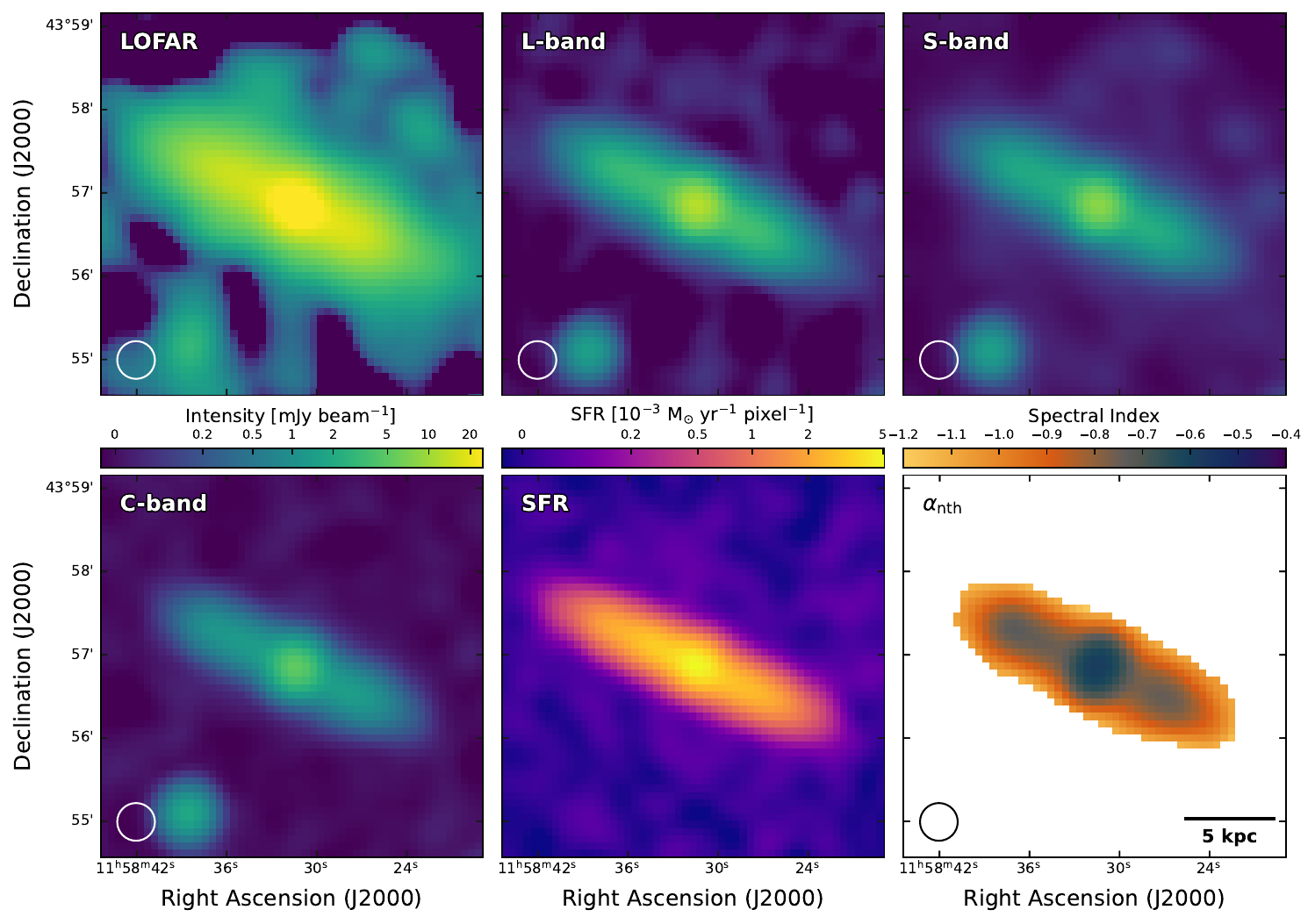}
    \caption{Multi-band maps of NGC~4013. See Figure~\ref{fig:exemplar_maps} for details.}
\end{figure*}

\begin{figure*}[ht!]
    \centering
    \includegraphics[width=0.82\linewidth]{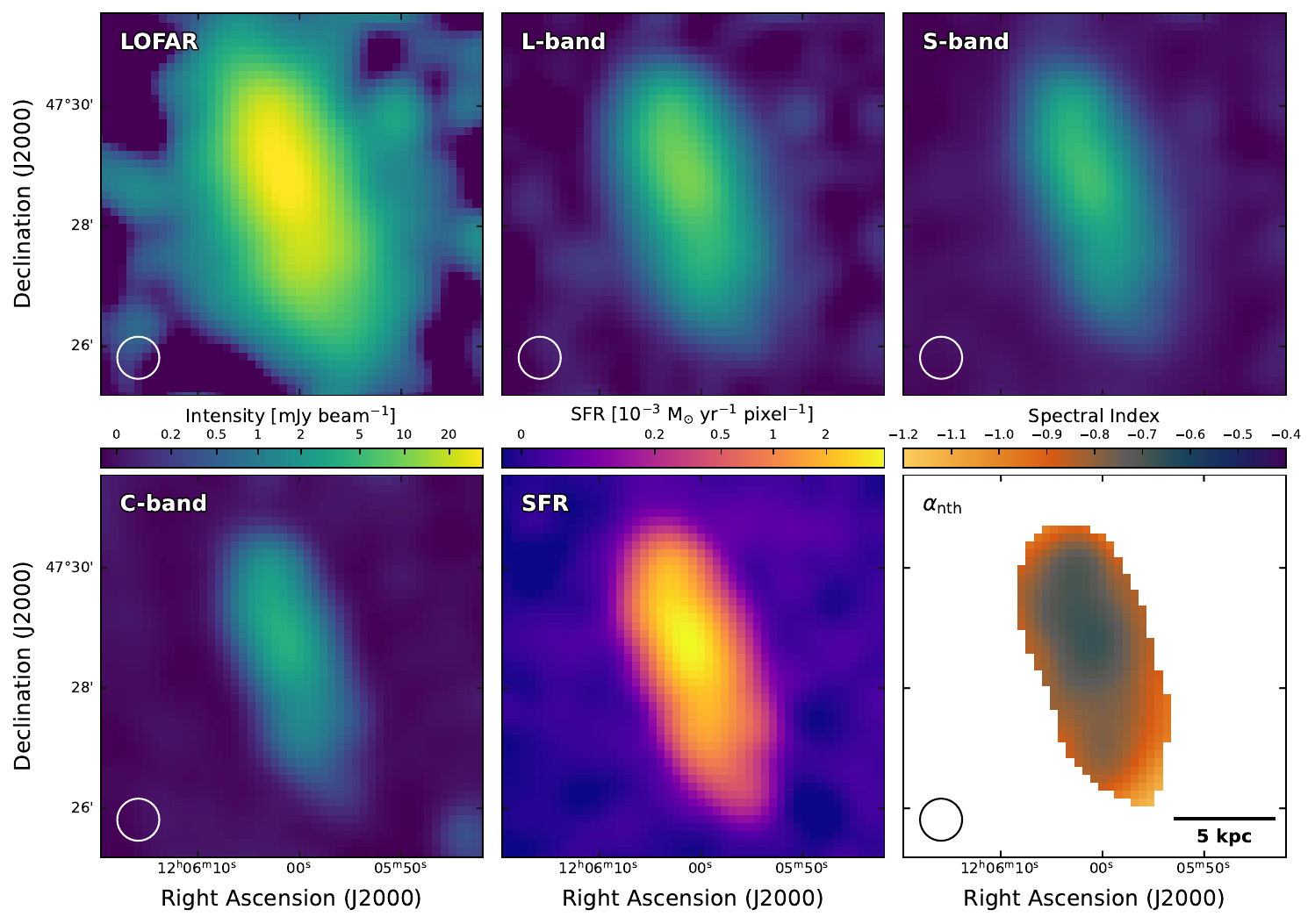}
    \caption{Multi-band maps of NGC~4096. See Figure~\ref{fig:exemplar_maps} for details.}
\end{figure*}

\begin{figure*}[ht!]
    \centering
    \includegraphics[width=0.82\linewidth]{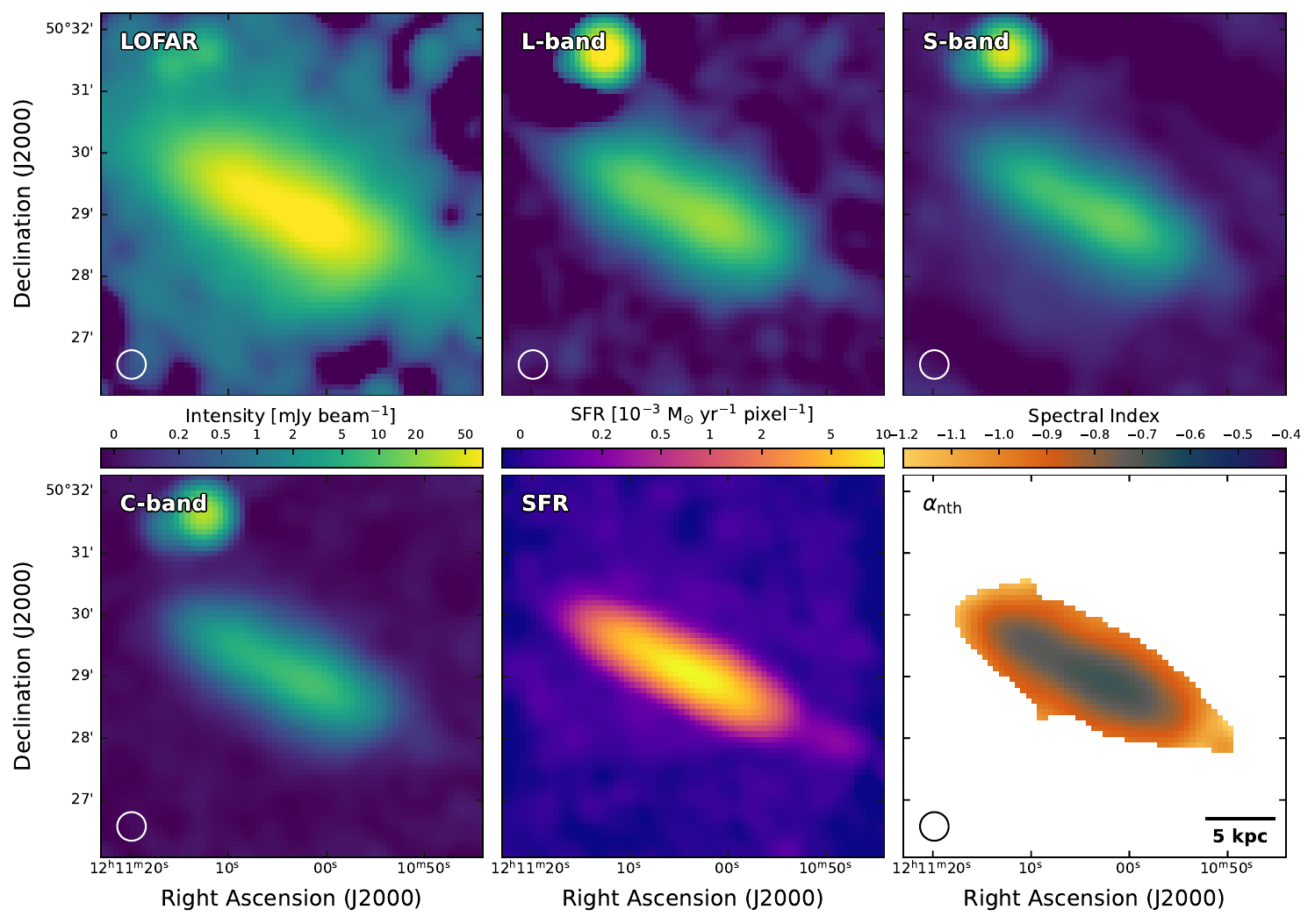}
    \caption{Multi-band maps of NGC~4157. See Figure~\ref{fig:exemplar_maps} for details.}
\end{figure*}

\begin{figure*}[ht!]
    \centering
    \includegraphics[width=0.82\linewidth]{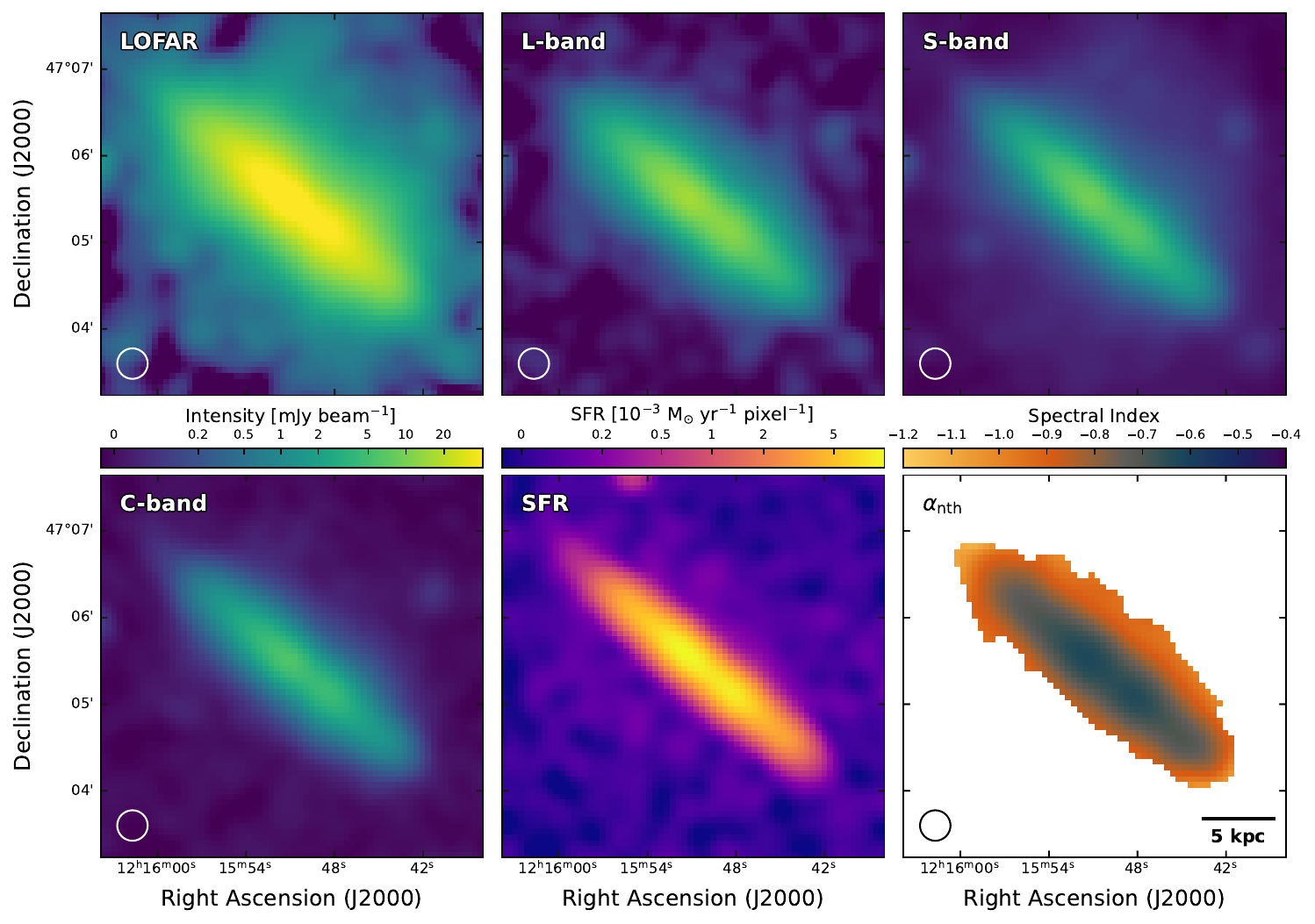}
    \caption{Multi-band maps of NGC~4217. See Figure~\ref{fig:exemplar_maps} for details.}
\end{figure*}

\begin{figure*}[ht!]
    \centering
    \includegraphics[width=0.82\linewidth]{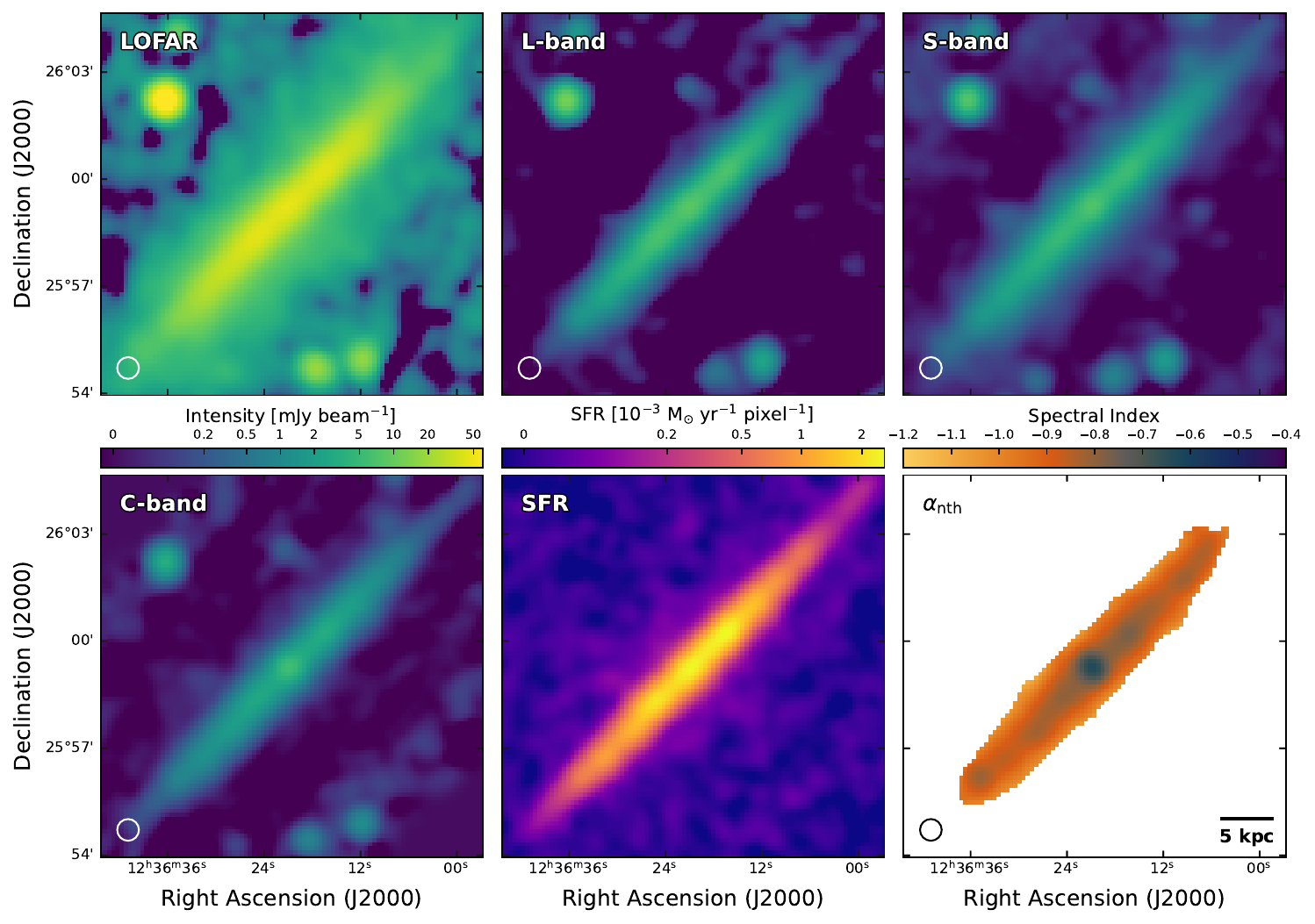}
    \caption{Multi-band maps of NGC~4565. See Figure~\ref{fig:exemplar_maps} for details.}
\end{figure*}

\begin{figure*}[ht!]
    \centering
    \includegraphics[width=0.82\linewidth]{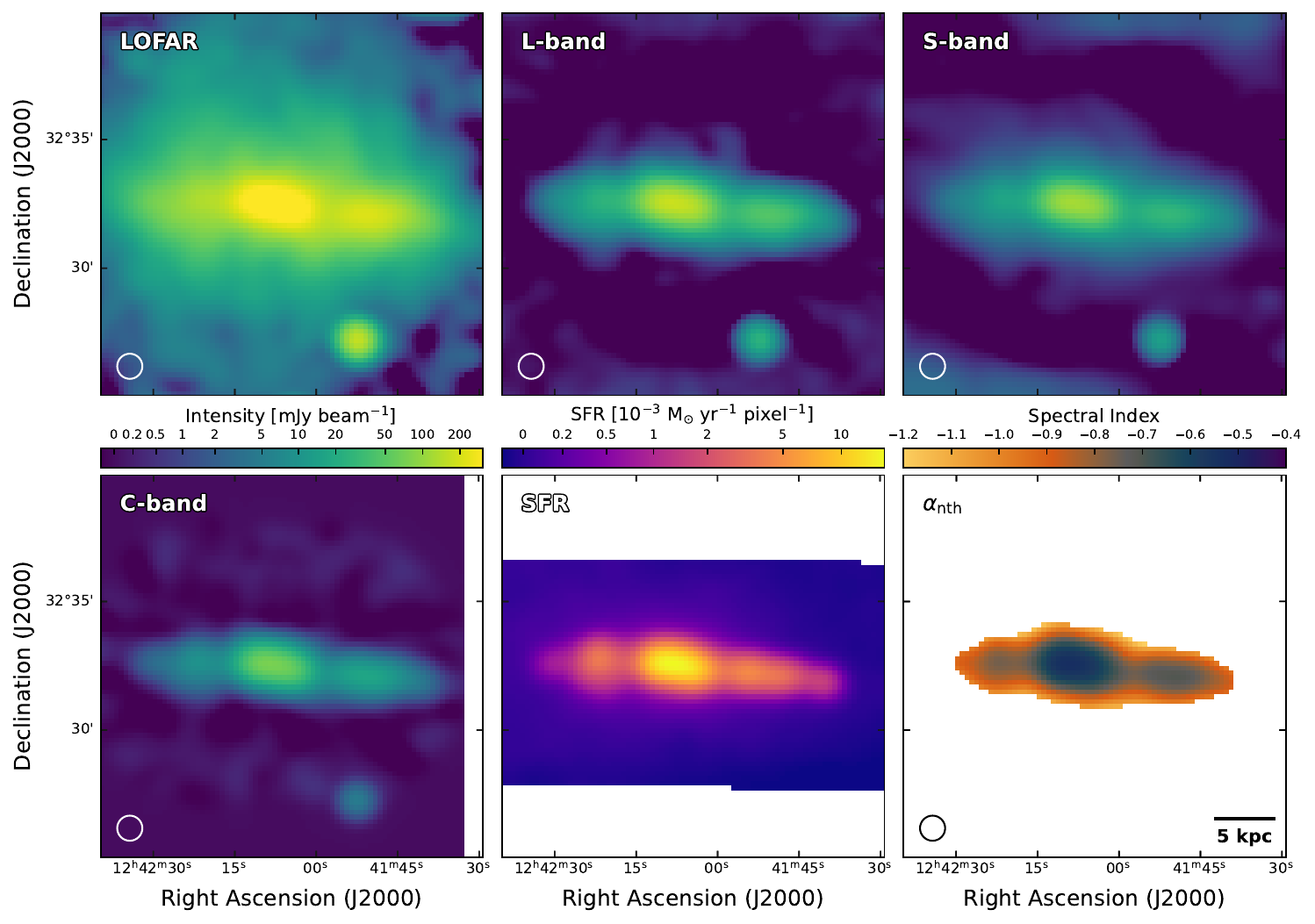}
    \caption{Multi-band maps of NGC~4631. See Figure~\ref{fig:exemplar_maps} for details.}
\end{figure*}

\begin{figure*}[ht!]
    \centering
    \includegraphics[width=0.82\linewidth]{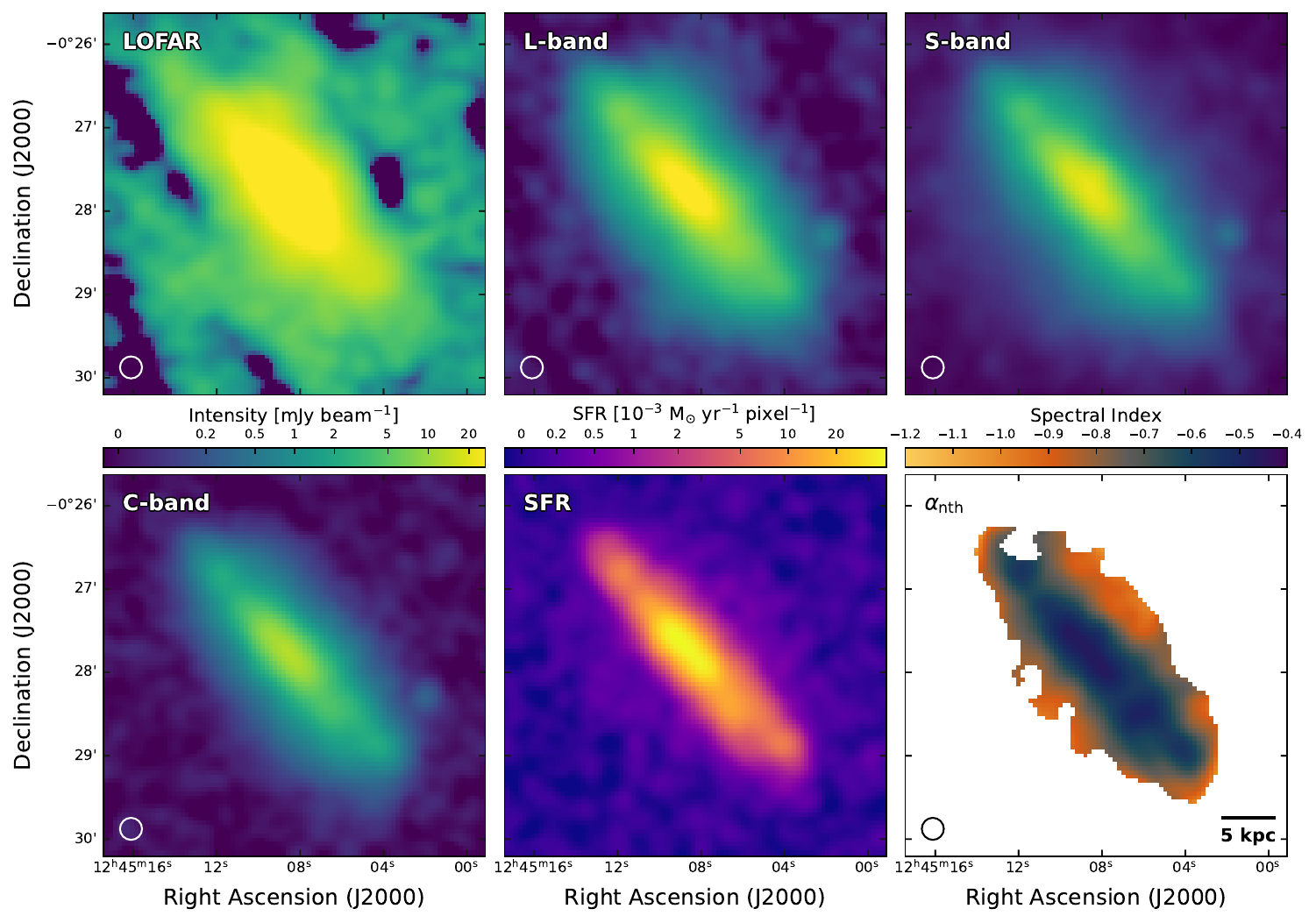}
    \caption{Multi-band maps of NGC~4666. See Figure~\ref{fig:exemplar_maps} for details.}
\end{figure*}

\begin{figure*}[ht!]
    \centering
    \includegraphics[width=0.82\linewidth]{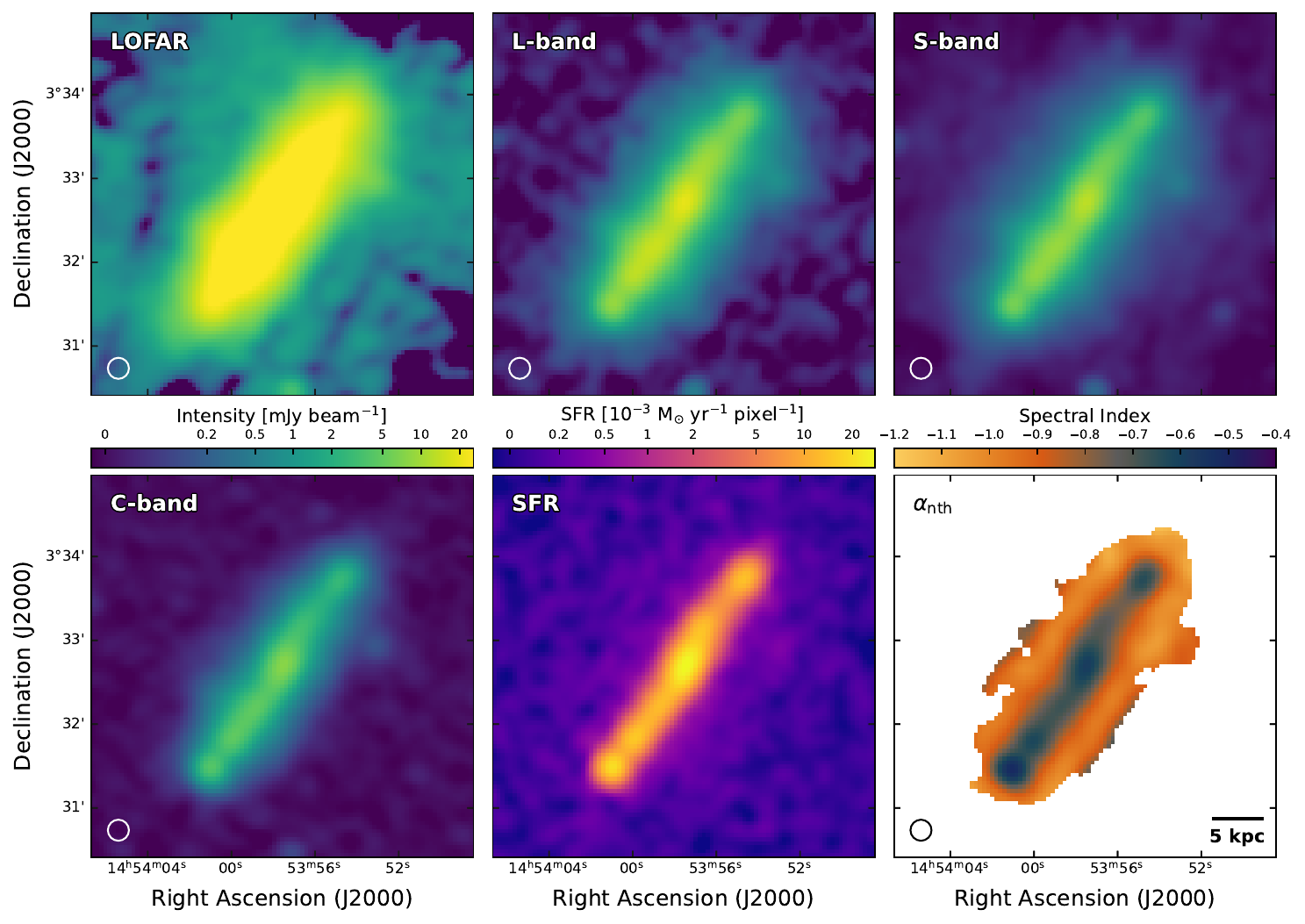}
    \caption{Multi-band maps of NGC~5775. See Figure~\ref{fig:exemplar_maps} for details.}
\end{figure*}

\begin{figure*}[ht!]
    \centering
    \includegraphics[width=0.82\linewidth]{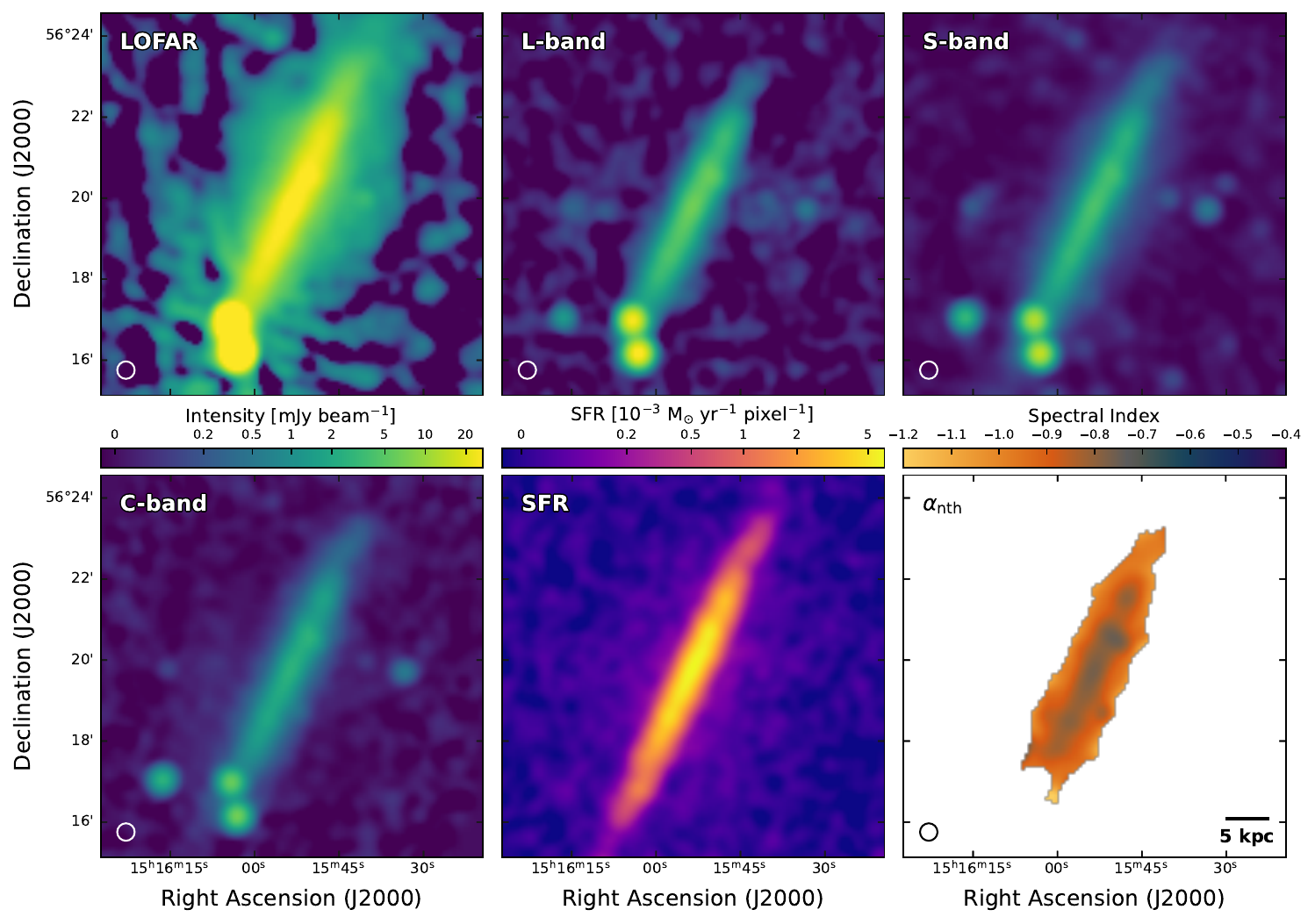}
    \caption{Multi-band maps of NGC~5907. See Figure~\ref{fig:exemplar_maps} for details.}
\end{figure*}

\begin{deluxetable*}{lcccccccc}
\tablecaption{Per-galaxy ODR Fitting Results for the Radio--SFR Relation
              \label{tab:rc_hyb_fits_pergal}}
\tablehead{
  \colhead{Galaxy} &
  \colhead{$k_{\rm LOFAR}$} & \colhead{$m_{\rm LOFAR}$} &
  \colhead{$k_L$} & \colhead{$m_L$} &
  \colhead{$k_S$} & \colhead{$m_S$} &
  \colhead{$k_C$} & \colhead{$m_C$}
}
\colnumbers
\tabletypesize{\footnotesize}
\startdata
NGC~891  & $0.63\pm0.04$ & $0.76\pm0.03$ & $1.05\pm0.06$ & $0.38\pm0.05$ & $0.94\pm0.05$ & $0.55\pm0.05$ & $1.19\pm0.05$ & $0.17\pm0.05$ \\
NGC~2683 & $0.53\pm0.03$ & $0.31\pm0.03$ & $0.76\pm0.05$ & $0.20\pm0.04$ & $0.73\pm0.03$ & $0.14\pm0.02$ & $0.91\pm0.08$ & $-0.15\pm0.07$ \\
NGC~2820 & $0.65\pm0.09$ & $0.77\pm0.06$ & $0.81\pm0.09$ & $0.78\pm0.06$ & $0.78\pm0.07$ & $0.73\pm0.05$ & $0.85\pm0.07$ & $0.63\pm0.06$ \\
NGC~3003 & $0.52\pm0.05$ & $0.48\pm0.04$ & $0.83\pm0.11$ & $0.34\pm0.08$ & $0.84\pm0.07$ & $0.24\pm0.06$ & $0.97\pm0.08$ & $0.12\pm0.07$ \\
NGC~3044 & $0.61\pm0.06$ & $0.62\pm0.06$ & $0.79\pm0.07$ & $0.69\pm0.06$ & $0.81\pm0.06$ & $0.66\pm0.06$ & $0.93\pm0.07$ & $0.49\pm0.07$ \\
NGC~3079 & $1.13\pm0.14$ & $0.82\pm0.11$ & $2.13\pm0.47$ & $0.19\pm0.30$ & $1.64\pm0.21$ & $0.36\pm0.16$ & $2.10\pm0.33$ & $0.02\pm0.24$ \\
NGC~3432 & $0.56\pm0.06$ & $0.35\pm0.05$ & $0.77\pm0.07$ & $0.32\pm0.06$ & $0.79\pm0.07$ & $0.27\pm0.06$ & $0.87\pm0.07$ & $0.10\pm0.07$ \\
NGC~3448 & $0.64\pm0.04$ & $0.44\pm0.04$ & $0.85\pm0.05$ & $0.43\pm0.05$ & $0.88\pm0.05$ & $0.36\pm0.05$ & $0.94\pm0.05$ & $0.30\pm0.06$ \\
NGC~3556 & $0.51\pm0.03$ & $0.66\pm0.03$ & $0.73\pm0.04$ & $0.49\pm0.04$ & $0.68\pm0.03$ & $0.51\pm0.03$ & $0.90\pm0.04$ & $0.16\pm0.05$ \\
NGC~3628 & $0.75\pm0.03$ & $0.49\pm0.03$ & $1.05\pm0.04$ & $0.35\pm0.05$ & $1.29\pm0.06$ & $-0.04\pm0.06$ & $1.34\pm0.05$ & $-0.09\pm0.05$ \\
NGC~4013 & $0.64\pm0.05$ & $0.38\pm0.04$ & $1.11\pm0.09$ & $0.00\pm0.07$ & $1.06\pm0.09$ & $0.06\pm0.08$ & $1.27\pm0.13$ & $-0.19\pm0.11$ \\
NGC~4096 & $0.62\pm0.03$ & $0.13\pm0.02$ & $0.88\pm0.06$ & $0.02\pm0.05$ & $0.84\pm0.06$ & $-0.08\pm0.04$ & $1.13\pm0.09$ & $-0.46\pm0.08$ \\
NGC~4157 & $0.52\pm0.03$ & $0.87\pm0.03$ & $0.69\pm0.04$ & $0.78\pm0.04$ & $0.71\pm0.04$ & $0.64\pm0.04$ & $0.76\pm0.04$ & $0.60\pm0.04$ \\
NGC~4217 & $0.56\pm0.04$ & $0.80\pm0.03$ & $0.73\pm0.06$ & $0.78\pm0.04$ & $0.77\pm0.05$ & $0.67\pm0.04$ & $0.87\pm0.06$ & $0.55\pm0.05$ \\
NGC~4565 & $0.53\pm0.02$ & $0.61\pm0.01$ & $1.00\pm0.04$ & $0.08\pm0.02$ & $0.74\pm0.03$ & $0.44\pm0.02$ & $0.96\pm0.03$ & $0.21\pm0.02$ \\
NGC~4631 & $0.52\pm0.03$ & $0.80\pm0.04$ & $0.90\pm0.05$ & $0.41\pm0.06$ & $0.84\pm0.05$ & $0.53\pm0.07$ & $1.17\pm0.06$ & $-0.03\pm0.08$ \\
NGC~4666 & $0.48\pm0.03$ & $1.02\pm0.03$ & $0.66\pm0.03$ & $1.07\pm0.04$ & $0.70\pm0.03$ & $0.96\pm0.04$ & $0.74\pm0.03$ & $0.86\pm0.04$ \\
NGC~5775 & $0.75\pm0.07$ & $1.03\pm0.05$ & $0.84\pm0.05$ & $0.90\pm0.05$ & $0.82\pm0.04$ & $0.87\pm0.04$ & $0.98\pm0.06$ & $0.59\pm0.05$ \\
NGC~5907 & $0.51\pm0.02$ & $0.61\pm0.02$ & $0.87\pm0.04$ & $0.13\pm0.04$ & $0.67\pm0.02$ & $0.38\pm0.02$ & $0.64\pm0.03$ & $0.39\pm0.02$ \\
\tableline
Median & $0.57^{+0.13}_{-0.07}$ & $0.63^{+0.22}_{-0.25}$ & $0.85^{+0.19}_{-0.13}$ & $0.40^{+0.39}_{-0.29}$ & $0.80^{+0.16}_{-0.09}$ & $0.44^{+0.25}_{-0.32}$ & $0.95^{+0.26}_{-0.15}$ & $0.20^{+0.38}_{-0.30}$ \\
\enddata
\tablecomments{Here, ``Radio'' and ``SFR'' refer to the nonthermal radio-derived and hybrid H$\alpha$+22~$\mu$m SFR surface densities, respectively, as defined in Section~\ref{subsec:sfr_sd}. Columns are as follows: (1) Galaxy designation. (2)--(3) ODR best-fit slope ($k$) and intercept ($m$) at 144~MHz (LOFAR). (4)--(5) Same at 1.575~GHz ($L$-band). (6)--(7) Same at 3.0~GHz ($S$-band). (8)--(9) Same at 6.0~GHz ($C$-band). All fits follow $\log_{10}\Sigma_{\rm SFR}^{\rm RC} = k\,\log_{10}\Sigma_{\rm SFR}^{\rm hyb} + m$ with surface densities in units of $10^{-3}\,M_{\odot}\,{\rm yr}^{-1}\,{\rm kpc}^{-2}$. The bottom row gives the sample medians with $16^{\rm th}$/$84^{\rm th}$ percentile uncertainties from bootstrap resampling.}
\end{deluxetable*}
\end{document}